\documentclass[aps,prd,onecolumn, twoside,notitlepage,preprintnumbers,showpacs,
showkeys,superscriptaddress,byrevtex,floatfix,nofootinbib,amsmath,amssymb,longbibliography]{revtex4-1}
\usepackage{xcolor}
\usepackage[T1]{fontenc}
\usepackage{mathrsfs}

\usepackage[thicklines,makeroom]{cancel}
\usepackage[all,color]{xy}
\usepackage{graphicx}% Include figure files
\usepackage{dcolumn}% Align table columns on decimal point
\usepackage{bm}% bold math
\usepackage{tensor}
\usepackage{array,amscd,amsmath,tikz, amssymb,amsthm,enumerate,mathtools,placeins,verbatim,tensor,multirow,dsfont}
\usepackage{rotating}
\usepackage[caption=false]{subfig}
\usepackage{eepic}
\usepackage[bookmarksnumbered,bookmarksopen=true,bookmarksopenlevel=1,pdfstartview=FitH,hidelinks]{hyperref}
\usepackage[all]{hypcap}

\newcommand{\R}{\mathds{R}}
\newcommand{\C}{\mathds{C}}
\newcommand{\Q}{\mathds{H}}
\newcommand{\Oct}{\mathds{O}}

\newcolumntype{L}[1]{>{\raggedright\let\newline\\\arraybackslash\hspace{0pt}}m{#1}}
\newcolumntype{C}[1]{>{\centering\let\newline\\\arraybackslash\hspace{0pt}}m{#1}}
\newcolumntype{R}[1]{>{\raggedleft\let\newline\\\arraybackslash\hspace{0pt}}m{#1}}
\newcolumntype{M}[1]{>{$}{#1}<{$}}

\newcommand{\aref}[1]{\hyperref[#1]{appendix~\ref{#1}}}

\def\0{{\sst{(0)}}}
\def\1{{\sst{(1)}}}
\def\2{{\sst{(2)}}}
\def\3{{\sst{(3)}}}
\def\4{{\sst{(4)}}}
\def\5{{\sst{(5)}}}
\def\6{{\sst{(6)}}}
\def\7{{\sst{(7)}}}

 \def\bd{\begin{document}} \def\ed{\end{document}}
\def\ds{\documentstyle} \let\fr=\frac \let\bl=\bigl \let\br=\bigr
\let\Br=\Bigr \let\Bl=\Bigl 
\let\bm=\bibitem
\let\na=\nabla
\let\pa=\partial \let\ov=\overline 
\newcommand{\be}{\begin{equation}} 
\newcommand{\ee}{\end{equation}} 
\def\ba{\begin{array}}
\def\ea{\end{array}}
\def\ft#1#2{{\textstyle{{\scriptstyle #1}\over {\scriptstyle #2}}}}
\def\fft#1#2{{#1 \over #2}}
\def\del{\partial}
\def\sst#1{{\scriptscriptstyle #1}}
\def\oneone{\rlap 1\mkern4mu{\rm l}}
\def\ie{{\it i.e.\ }}
\def\via{{\it via}}
\def\semi{{\ltimes}}
\def\v{{\cal V}}

\newcommand{\bp}{\bullet}
\newcommand{\sfx}{\textsf{x}}
\newcommand{\sfo}{\textsf{o}}
\newcommand{\alg}[1]{\ensuremath{\mathfrak{#1}}}
\newcommand{\rep}[1]{\ensuremath{\mathbf{#1}}}
\newcommand{\fld}[1]{\ensuremath{\mathds{#1}}}
\newcommand{\rng}[1]{\ensuremath{\mathds{#1}}}
\newcommand{\SUSY}{\ensuremath{\mathcal{N}}}
\newcommand{\N}{\ensuremath{\mathcal{N}}}
\newcommand{\susy}{\ensuremath{\mathcal{N}}}
\newcommand{\tN}{\ensuremath{\tilde{\mathcal{N}}}}
\newcommand{\J}{\mathfrak{J}}

\newcommand{\tbV}{\tilde{\mathbf{V}}}
\newcommand{\tbC}{\tilde{\mathbf{C}}}
\newcommand{\tA}{\tilde{A}}
\newcommand{\tX}{\tilde{\chi}}
\newcommand{\tphi}{\tilde{\phi}}
\newcommand{\tbG}{\tilde{\mathbf{G}}}
\newcommand{\tG}{\tilde{G}}

\newcommand{\bV}{\mathbf{V}}
\newcommand{\bC}{\mathbf{C}}
\newcommand{\bM}{\mathbf{M}}
\newcommand{\bH}{\mathbf{H}}
\newcommand{\bG}{\mathbf{G}}
\newcommand{\bT}{\mathbf{T}}
\newcommand{\tbT}{\tilde{\mathbf{T}}}

\DeclareMathOperator{\Aut}{Aut}
\DeclareMathOperator{\Str}{Str}
\DeclareMathOperator{\tr}{tr}
\DeclareMathOperator{\Tr}{Tr} % for partial traces only
\DeclareMathOperator{\Det}{Det}
\DeclareMathOperator{\Pf}{Pf}
\DeclareMathOperator{\diag}{diag}
\DeclareMathOperator{\Iso}{Iso}
\DeclareMathOperator{\Hom}{Hom}
\DeclareMathOperator{\End}{End}
\DeclareMathOperator{\SO}{SO}
\DeclareMathOperator{\Orth}{O}
\DeclareMathOperator{\USp}{USp}
\DeclareMathOperator{\SL}{SL}
\DeclareMathOperator{\SU}{SU}
\DeclareMathOperator{\Sp}{Sp}
\DeclareMathOperator{\ISO}{ISO}
\DeclareMathOperator{\Un}{U}
\DeclareMathOperator{\Span}{span}
\DeclareMathOperator{\Cliff}{Cliff}
\DeclareMathOperator{\Pin}{Pin}
\DeclareMathOperator{\Spin}{Spin}

\newcommand{\al}{\mathds{A}}
\newcommand{\Alg}{\mathds{A}}
\newcommand{\mf}{\mathfrak}
\newcommand{\F}{\mathds{F}}

\newcommand{\nl}{{\mathcal{N}}}
\newcommand{\nr}{{\tilde{\mathcal{N}}}}

\newcommand{\ho}[1]{$\, ^{#1}$}
\newcommand{\hoch}[1]{$\, ^{#1}$}
\newcommand{\bea}{\begin{eqnarray}} 
\newcommand{\eea}{\end{eqnarray}} 
\newcommand{\ra}{\rightarrow}
\newcommand{\lra}{\longrightarrow}
\newcommand{\Lra}{\Leftrightarrow}

\usepackage{xparse}

\makeatletter
\NewDocumentCommand{\raisedminus}{m}{%
  \raisebox{0.2em}{$\m@th#1{-}$}%
}
\NewDocumentCommand{\unaryminus}{}{%
  \mathbin{%
    \mathchoice{%
      \raisedminus\scriptstyle
    }{%
      \raisedminus\scriptstyle
    }{%
      \raisedminus\scriptscriptstyle
    }{%
      \raisedminus\scriptscriptstyle
    }%
  }%
}
\makeatother

\begin{document}

\preprint{DIAS-STP-16-08}\preprint{Imperial/TP/2016/MJD/02}
\preprint{DFPD/2016/TH-07}

\title{Twin Supergravities from Yang-Mills Squared} 

\author{A. Anastasiou}
\email[]{alexandros.anastasiousu@su.se}
\affiliation{Nordita, KTH Royal Institute of Technology and Stockholm University, Roslagstullsbacken 23, 10691 Stockholm, Sweden}
\author{L. Borsten}
\email[]{leron@stp.dias.ie}
\affiliation{School of Theoretical Physics, Dublin Institute for Advanced Studies,
10 Burlington Road, Dublin 4, Ireland}
\author{M. J. Duff}
\email[]{m.duff@imperial.ac.uk}
\affiliation{Theoretical Physics, Blackett Laboratory, Imperial College London,
London SW7 2AZ, United Kingdom}
\author{M. J. Hughes}
\email[]{mia.hughes07@imperial.ac.uk}
\affiliation{Theoretical Physics, Blackett Laboratory, Imperial College London,
London SW7 2AZ, United Kingdom}
\author{A. Marrani}
\email[]{alessio.marrani@pd.infn.it}
\affiliation{Centro Studi e Ricerche ``Enrico Fermi'',
Via Panisperna 89A, I-00184, Roma, Italy}
\affiliation{Dipartimento di Fisica e Astronomia ``Galileo Galilei'', Universit\`a di Padova, and INFN, sezione di
Padova,
Via Marzolo 8, I-35131 Padova, Italy}
\author{S. Nagy}
\email[]{snagy@math.tecnico.ulisboa.pt}
\affiliation{Centro de An\'alise Matem\'atica, Geometria e Sistemas Din\^amicos
Pavilh\~ao de Matem\'atica
Instituto Superior T\'ecnico
Av.~Rovisco Pais
1049-001 Lisboa
Portugal}
\author{M. Zoccali}
\email[]{m.zoccali14@imperial.ac.uk}
\affiliation{Theoretical Physics, Blackett Laboratory, Imperial College London,
London SW7 2AZ, United Kingdom}
\date{\today}

\begin{abstract}
We consider  `twin supergravities' - pairs of supergravities with $\N_+$ and $\N_-$ supersymmetries, $\N_+>\N_-$, with identical bosonic sectors - in the context of tensoring super Yang-Mills multiplets. It is demonstrated that the pairs of twin supergravity theories are related through their left and right super Yang-Mills factors.  This procedure generates new theories from old. In particular, the matter coupled $\N_-$ twins in $D=3,5,6$ and the $\N_-=1$ twins in $D=4$ have not, as far as we are aware, been obtained previously using the double-copy construction, adding to the growing list of double-copy constructible theories. The use of fundamental matter multiplets in the double-copy construction leads us to introduce a bi-fundamental scalar that couples to the well-known bi-adjoint scalar field. It is also shown that certain matter coupled supergravities admit more than one factorisation into left and right super Yang-Mills-matter theories.
% A classification of all possible factorisations under the assumption that the  scalar manifold of the supergravity theory is homogeneous is given.
\end{abstract}

\pacs{11.30.Pb,11.15.-q,02.20.-a,02.20.-a}
\keywords{Yang-Mills theory, supergravity, amplitudes, symmetries}

\maketitle
\tableofcontents

\newpage

\section{Introduction}

It has been known for some time \cite{Gunaydin:1983rk, Dolivet:2007sz, Bianchi:2007va} that there exist pairs of supergravity theories  with identical bosonic  sectors, both in terms of content and couplings, but distinct degrees of supersymmetry $\N_+>\N_-$. The canonical example in $D=4$ dimensions is given by  $\N_+=6$ supergravity and the magic $\N_-=2$  supergravity coupled to 15 vector multiplets, which  despite their  distinct supersymmetric completions  have the same bosonic Lagrangian with scalar coset $\SO^\star(12)/\Un(6)$. All such `twin' supergravities, which we will denote  by $(\N_+, \N_-)$,  were classified in \cite{Roest:2009sn,Duff:2010ss}.

A so far unrelated idea is that of gravity as `the square of gauge theory'. Schematically\footnote{The precise meaning of this product and the issue of the gauge indices is discussed in \cite{Anastasiou:2014qba, Borsten:2015pla} and given in \eqref{offshell}.}, 
\be
A_\mu \otimes \tilde{A}_\nu = g_{\mu\nu} \oplus B_{\mu\nu} \oplus \varphi .
\ee
Here, $A_\mu$ and $\tilde{A}_\nu$ are the gauge potentials of two distinct Yang-Mills theories, which we will refer to as left (no tilde) and right (tilde), respectively. They can have arbitrary and independent  non-Abelian gauge groups $G$ and $\tG$. Beyond the identification of asymptotic on-mass-shell states, this formal identity can be motivated by the Kawai-Lewellen-Tye (KLT) relations, which connect tree-level amplitudes of closed strings to sums of products of open string amplitudes \cite{Kawai:1985xq}. More recently,  invoking Bern-Carrasco-Johansson  (BCJ) colour-kinematic duality \cite{Bern:2008qj} it has been  conjectured  \cite{Bern:2010yg,Bern:2010ue} that the on-mass-shell momentum-space scattering amplitudes for gravity are the  ``double-copy''  of gluon scattering amplitudes in Yang-Mills theory to all orders in perturbation theory. These relations have since been generalised to a large class of (super)gravity theories, including a variety of matter couplings and even pure gravity \cite{Carrasco:2012ca, Damgaard:2012fb, Huang:2012wr, Bargheer:2012gv, Johansson:2014zca, Chiodaroli:2014xia, Chiodaroli:2015rdg, Chiodaroli:2015wal, Chiodaroli:2016jqw}. For reviews see \cite{Elvang:2013cua,Carrasco:2015iwa}. The double-copy prescription has proven itself a tremendously effective computational tool, pushing the boundaries of what can be achieved in perturbative quantum gravity and in the process revealing numerous surprises \cite{Katsaroumpas:2009iy, Bern:2009kd, Bern:2013uka, Carrasco:2013ypa, Bern:2013yya, Bern:2013qca,Bern:2014lha, Bern:2014sna, Bern:2015xsa}. There is now a growing literature \cite{Bern:2014kca,BjerrumBohr:2010rt,Bern:2015ooa, Cachazo:2013iea, Dolan:2013isa,Cachazo:2014xea,Naculich:2014naa, Naculich:2015zha, Du:2016tbc, Du:2016blz, Nandan:2016pya, Bjerrum-Bohr:2016axv,Monteiro:2011pc,BjerrumBohr:2012mg,Monteiro:2013rya,Boels:2013bi,Fu:2016plh,BjerrumBohr:2009rd,Stieberger:2009hq,Mafra:2011kj,Broedel:2012rc,Broedel:2013tta,Mafra:2014gja,Mafra:2015vca,Mafra:2015mja,He:2015wgf, Schlotterer:2016cxa, Carrasco:2016ldy,Mason:2013sva, Geyer:2014fka,Casali:2015vta,Geyer:2015bja,Geyer:2015jch,Geyer:2016wjx,Chiodaroli:2011pp, Borsten:2013bp, Anastasiou:2013cya, Anastasiou:2013hba, Anastasiou:2014qba, Nagy:2014jza, Anastasiou:2015vba, Borsten:2015pla,Boels:2016xhc,Monteiro:2014cda, Luna:2015paa, Luna:2016due, White:2016jzc, Cardoso:2016ngt} expanding upon, and refining  our understanding of, these remarkable relations, which has benefitted from a  diverse array of complementary  approaches: scattering equations \cite{Cachazo:2013iea, Dolan:2013isa,Cachazo:2014xea,Naculich:2014naa, Naculich:2015zha, Du:2016tbc, Du:2016blz, Nandan:2016pya, Bjerrum-Bohr:2016axv},  kinematic algebras \cite{Monteiro:2011pc,BjerrumBohr:2012mg,Monteiro:2013rya,Boels:2013bi,Fu:2016plh}, string theory \cite{BjerrumBohr:2009rd,Stieberger:2009hq,Mafra:2011kj,Broedel:2012rc,Broedel:2013tta,Mafra:2014gja,Mafra:2015vca,Mafra:2015mja,He:2015wgf, Schlotterer:2016cxa, Carrasco:2016ldy},  twistor theory \cite{Mason:2013sva, Geyer:2014fka,Casali:2015vta,Geyer:2015bja,Geyer:2015jch,Geyer:2016wjx}, on-shell and off-shell symmetries \cite{Chiodaroli:2011pp, Borsten:2013bp, Anastasiou:2013cya, Anastasiou:2013hba, Anastasiou:2014qba, Nagy:2014jza, Anastasiou:2015vba, Borsten:2015pla}, minimal physical assumptions \cite{Boels:2016xhc} and  even non-perturbative classical solutions \cite{Monteiro:2014cda, Luna:2015paa, Luna:2016due, White:2016jzc, Cardoso:2016ngt}.

  In the present paper we look at  twin supergravities in this context, studying the relationship between the two Yang-Mills factors generating   the twin supergravity theories.  In \cite{Borsten:2013bp, Anastasiou:2013hba, Anastasiou:2015vba} all possible  products of two super Yang-Mills theories, with \emph{no} additional matter couplings, in  dimensions $3\leq D \leq10$ were considered yielding a pyramid of supergravity theories and  their corresponding scalar manifolds. See \autoref{PYR}. Remarkably, all theories appearing in the pyramid, with the exception of the maximal ``spine'' indicated in red, have a twin with fewer supersymmetries.

  It is demonstrated here that such twin supergravity theories are related in a controlled manner through their Yang-Mills factors.  Each twin pair $(\N_+, \N_-)$ can be regarded as a pair of complementary consistent truncations of a single $(\mathscr{N}=\N_++\N_-)$-extended parent  supergravity theory  \cite{Roest:2009sn}, which is given by the product of a left $\nl$-extended Yang-Mills theory with a right $\tilde{\N}$-extended Yang-Mills theory,  where $\mathscr{N}=\nl+\nr$ and without loss of generality we consider $\nl\geq\nr$. The twins relations are mediated by the Yang-Mills factors as depicted schematically here:
\be
    \xymatrix{
             & {\begin{array}{c}\text{Parent supergravity}\\ \bG_{\nl+\nr}\oplus\bM_{\nl+\nr}\end{array}}  \ar[d]^{\text{Yang-Mills factors}} &   \\
         & \bV_\nl \otimes \tbV_\nr \ar[dl] \ar[dr]  &   \\
         [\bV_{\nl'}\oplus\bC_{\nl'}^{\rho}] \otimes \tbV_\nr   \ar[d]   \ar@{<-->}[rr]_{\text{twin relation}}               &  &  [\bV_{\nl'}\oplus\bC_{\nl'}^{\rho}] \otimes [\tbV_{\nr'}\oplus\tbC_{\nr'}^{\tilde{\rho}}]  \ar[d] \\{\begin{array}{c}\N_+ ~ \text{big twin supergravity}\\ 
     \bG_{\N_+}\oplus\bM_{\N_+}  \end{array}}    &  &{\begin{array}{c}\N_- ~ \text{little twin supergravity}\\  \bG_{\N_-}\oplus\bM_{\N_-} \end{array}}  }
\ee
Here,  $\bG_\N, \bV_\N, \bC_\N$ and $\bM_\N$ denote   $\N$-extended gravity, vector, spinor  and generic (not necessarily irreducible) matter multiplets, respectively.  The left $\bV_\nl$ and  right $\tbV_\nr$ multiplets of the parent supergravity are decomposed into $\nl'<\nl$ and $\nr'<\nr$ multiplets, $\bV_{\nl'}\oplus\bC_{\nl'}^{\rho}\oplus\cdots$ and $\tbV_{\nr'}\oplus\tbC_{\nr'}^{\tilde{\rho}}\oplus\cdots$, and  the resulting adjoint spinor multiplets are replaced by fundamental\footnote{We use fundamental here loosely to refer to gauge group representations other than the adjoint.}  multiplets as indicated by the superscript $\rho$. This procedure generates new theories from old. In particular, the matter coupled $\N_-$ twins in $D=3,5,6$ and the $\N_-=1$ twins in $D=4$ have not, as far as we are aware, been obtained previously using the double-copy construction, adding to the growing list of double-copy constructible theories. The twin theories in $D=3,4,5,6$ and their left/right (super) Yang-Mills factorisations as determined by the above prescription are given in \autoref{D3},  \autoref{D4},  \autoref{D5} and  \autoref{D6}. The use of fundamental matter multiplets in the double-copy construction leads us to introduce a bi-fundamental scalar that couples to the well-known bi-adjoint scalar field. It is also shown that certain matter coupled supergravities admit more than one factorisation into left and right super Yang-Mills-matter theories.

The remaining sections are organised as follows. In \autoref{twins} we review the classification of all twin supergravities. In \autoref{twinsquare} we demonstrate  how the pyramid twins are related via  Yang-Mills squared. We outline the general procedure and discuss the bi-fundamental scalar theory before presenting a detailed example in \autoref{example}. In \autoref{summary} we summarise the pyramid of twins and make some additional comments on the $D=6$ and $D=3$ cases. The \emph{triplets} are considered \autoref{triplets}. In \autoref{othertwins} we treat the isolated twin pair not appearing in the pyramid and the generalisation of the $D=4, (2,1)$ twin pair to a sequence of $D=4, (2,1)$ twin pairs. We conclude in \autoref{conclusions} with a summary and future directions. 
%In \autoref{alternative} we present a  classification of all possible factorisations of supergravity theories into Yang-Mills-matters under the assumption that the  scalar manifold of the supergravity theory is homogeneous.

\section{Twin supergravity theories}\label{twins}

 Twin supergravities are theories that have identical bosonic Lagrangians, but different supersymmetric completions. Twin pairs are denoted by $(\susy_+,\susy_-)$ with $\susy_+>\susy_-$.  Such theories appear in $D=3,4,5,6$. In $D=3$ where all vectors dualise to scalars, matching of the scalar cosets is a sufficient condition for twinness. However, this criterion is necessary but not sufficient in $D=4,5,6$.

Here we summarise the classification of twin supergravities provided
in \cite{Roest:2009sn,Duff:2010ss}. The classification is done in $D=3$ and relies  on listing the scalar manifolds of theories with different $\susy$ and then checking whether any of them match. The scalar manifolds of $D=3$ supergravities are given in  \autoref{MSCALAR3} where we distinguish between the theories with K\"ahler and Quaternionic manifolds, the matter-coupled theories and the unique pure supergravity theories. \begin{table}[h!]
$\begin{array}{crc}
\toprule
&&\\
~~\susy~~ & & \mathcal{M}_{scalar(3)}\\
&&\\
\hline
\hline
&&\\
1 & & \text{Riemannian}\\[5pt]
2 & & \text{K\"ahler}\\[5pt]
3 & & \text{Quaternionic}\\[5pt]
4 & & \text{Quaternionic $\times$ Quaternionic}\\[5pt]
&&\\
5 & & \frac{\USp(1,n)}{\Un(1)\times{\USp(n)}}\\[5pt]
6 & & \frac{\SU(4,n)}{\SU(4)\times{\SU(n)}\times{\Un(1)}}\\[5pt]
8 & & \frac{\SO(8,n)}{\SO(8)\times{\SO(n)}}\\[5pt]
&&\\
9 & & \frac{F_{4(-20)}}{\SO(9)}\\[5pt]
10 & & \frac{E_{6(-14)}}{\SO(10)\times{\SO(2)}}\\[5pt]
12 & & \frac{E_{7(-5)}}{\SO(12)\times{\SO(3)}}\\[5pt]
16 & & \frac{E_{8(8)}}{\SO(16)}\\[5pt]
&&\\
\hline
\hline
\end{array}$
\label{MSCALAR3}
\caption{The scalar manifolds of $D=3$ supergravity theories with $\susy$
supercharges.}
\end{table}
All scalar manifolds in  \autoref{MSCALAR3} are Riemannian
and therefore all theories could be thought as $\susy=1$ theories. 
Such twins are in this sense trivial. There are, however, a number that are particularly natural from the perspective of Yang-Mills squared, which we will therefore include in our analysis. Furthermore, all $\susy=3$ theories can be interpreted as $\susy=4$ with a trivial second quaternionic factor and thus are omitted in the analysis that follows. In $D=3$, where all vectors are dual to scalars, two theories that have the same scalar manifold are  twins. To carry out the classification one needs to check whether any of the scalar manifolds for $\susy\geq5$ are K\"ahler, quaternionic or both. A list of the possible K\"ahler and Quaternionic manifolds is provided in  \aref{MANIFOLDS}. A matching of the scalar manifolds gives the classification provided in \cite{Roest:2009sn}. 
\begin{table}[h!]
$\begin{array}{crcrc}
\toprule
&&\\
~~(\susy_+,\susy_-)~~ & & \mathcal{M}_{scalar} & & ~~D_{max}~~\\
&&\\
\hline
\hline
&&\\
(4,2) & & \frac{\SU(2,p)}{\SU(2)\times{\Un(p)}}\times\frac{\SU(2,q)}{\SU(2)\times{\Un(q)}} & & 4\\
&&\\
(6,2) & & \frac{\SU(4,p)}{\SU(4)\times{\Un(p)}} & & \hspace{4pt}4^*\\
&&\\
(8,2) & & \frac{\SO(8,2)}{\SO(8)\times{\SO(2)}} & & 4\\
&&\\
(10,2) & & \frac{E_{6(-14)}}{\SO(10)\times{\SO(2)}} & & 4\\
&&\\
(5,4) & & \frac{\USp(2,1)}{\USp(2)\times{\Un(1)}} & & 3\\
&&\\
(8,4) & & \frac{\SO(8,4)}{\SO(8)\times{\SO(4)}} & & 6\\
&&\\
(12,4) & & \frac{E_{7(-5)}}{\SO(12)\times{\SO(3)}} & & 6\\
&&\\
\hline
\hline
\end{array}$
\label{TWINS3}
\caption{The twin supergravity theories in $D=3$. $D_{max}$ is the highest
dimension to which these theories can be uplifted. \hspace{2cm}$^*$ The (6,2) sequence admits an uplift to D=4 only for $p=2$, since for all other values of $p$ it oxidises to $\susy=3$, $D=4$ theories whose kinetic vector matrix in non-holomorphic, which cannot be twins to $\susy=1$. This refines the treatment in \cite{Roest:2009sn}.}
\end{table}

The classification provided in  \autoref{TWINS3} differs from the one in \cite{Roest:2009sn} in the $(4,2)$ entry where the authors give only one of the two factors. Since the product of two K\"ahler manifolds is K\"ahler  the second factor is allowed. As the authors mention, it is clear from  \autoref{TWINS3} that the theory with scalar manifold $\SU(4,2)/\SU(4)\times{\SU(2)}\times{\Un(1)}$ has three supersymmetric completions and we refer to it as the triplet $(6,4,2)$. More generally, triplets will be denoted $(\N_+, \N_{-}^{+}, \N_{-}^{-})$, where  $\N_+$ is the big sibling of both $\N^{\pm}_{-}$. The two cases, $(\N_+, \N^{+}_{-})$ and $(\N_+, \N^{-}_{-})$, are conventional twin pairs in the sense that they follow the same pattern as the pure twins, as described in \autoref{triplets}.

 All twin pairs in higher dimensions can be obtained by oxidation of the $D=3$ ones \cite{Roest:2009sn}. We now  oxidise the theories from $D=3\rightarrow4$ (halves $\susy$),  $D=4\rightarrow5$ (preserves $\susy$) and  $D=5\rightarrow6$ (halves $\susy$) to obtain the twin pairs together with their scalar manifolds in the higher dimensional theories. We do so to demonstrate the crucial point that \textit{although all twin pairs can be obtained from oxidation, not all oxidised pairs form twins}. Matching bosonic content and scalar manifolds is a necessary but not sufficient condition in $D>3$.  A simple example illustrating this point in $D=4$ is given by the scalar manifold $\SU(3,3)/[\Un(3)\times{\SU(3)}]$, which  occurs three times: $(i)$ in $\susy=1$, $(ii)$ in $\susy=3$ and $(iii)$ in $\susy=2$. These theories have $64$, $64$, $80$ degrees of freedom respectively and thus clearly the latter cannot form a twin pair with either of the other two. The $\susy=3$ theory is coupled to three vector multiplets and the kinetic vector multiplet is non-holomorphic. Since all $\susy=1$ supergravities in $D=4$ must have a holomorphic kinetic vector matrix, the first two theories cannot form a twin pair either.

A second example in $D=4$ that serves to highlight the various subtleties is  given by  the coset $\SU(1,1)/\Un(1)$. For $\mathcal{N}\geqslant 2$ the scalar
manifold $\SU(1,1)/\Un(1)$ occurs three times:
\begin{enumerate}
\item  $\mathcal{N}=4$ pure supergravity \cite{Cremmer:1977tt}: the U-duality group is $\Un(4)\simeq \SO(6)\times \SU(1,1)$, with  scalar
manifold,
\be
\SU(1,1)/\Un(1)\times \SO(6)/\SO(6)\cong\SU(1,1)/\Un(1).
\ee As
such, this is the $n=0$ element of the  sequence $\SU(1,1)/\Un(1)\times
\SO(6,n)/\left[ \SO(6)\times \SO(n)\right] $, where $n$ denotes the number of vector
multiplets coupled to the $\mathcal{N}=4$ gravity multiplet. The Abelian $2$-form field strengths and their duals sit in the $\left( \mathbf{6},\mathbf{2}\right) $ of $\SO(6)\times \SU(1,1)$.   The pair $\left( \SO(6)\times \SU(1,1),\left( \mathbf{6},%
\mathbf{2}\right) \right) $ defines a ``group  of type $E_{7}$'' \cite{Brown:1969} of a very particular character \cite{Ferrara:2008ap}: while in
the bare charges basis, the invariant is quartic, it becomes a perfect
square in the dressed (supersymmetry) charges basis\footnote{%
The only other known example of a group  of type $E_{7}$
sharing this property is the pair $\left( \SU(5,1),\mathbf{20}\right) $ of $%
\mathcal{N}=5$, $D=4$ supergravity \cite{Ferrara:2008ap}.}.

\item  $\N=2$ supergravity \textit{minimally coupled} to a single vector multiplet
 (\emph{i.e.}~the  $\mathcal{N}=2$ axion-dilaton
model) \cite{Luciani:1977hp}: the U-duality group is  $\Un(1,1)$ with
scalar manifold, 
\be
\overline{\mathbb{CP}}^{1}\cong
\SU(1,1)/\Un(1)\times \Un(1)/\Un(1)\cong \SU(1,1)/\Un(1).\ee
 As such, this is the $n=1$ element of the
 sequence $\overline{\mathbb{CP}}^{n}\cong \Un(1,n)/\left[ \Un(1)\times
\Un(n)\right] $, where $n$ denotes the number of vector multiplets minimally coupled to
the $\mathcal{N}=2$ gravity multiplet. The Abelian $2$-form field strengths
and their duals sit in the $\mathbf{2}_{1}+\mathbf{2}_{-1}$ of $\Un(1,1)$. The global $\Un(1)$ factor is a relic of the compact symmetry of the
Maxwell theory of the lone graviphoton to which the electromagnetic sector reduces if the vector multiplet is
truncated. Note that
the axion-dilaton model is a consistent truncation of the $\mathcal{N}=4$
pure supergravity considered at point 1. In the bosonic sector this amounts to
removing four graviphotons out of six; in this way, the $\mathcal{N}=2$
axion-dilaton model is obtained in a symplectic frame in which the
holomorphic prepotential reads $F=-iX^{0}X^{1}$, as opposed to the  manifestly $\SU(1,1)$-symmetric Fubini-Study symplectic frame for which, 
\be
F=-\frac{i}{2}\left[ \left( X^{0}\right) ^{2}-\left( X^{1}\right) ^{2}
\right].\ee
 The two frames are related by a global $\USp\left( 4, \R
\right) $ transformation \cite{Ferrara:2010cw}. The pair $\left( \Un(1,1),\mathbf{2}_{1}+\mathbf{2}_{-1}\right) $ defines a \textit{degenerate} group
 of type $E_{7}$ \cite{Ferrara:2012qp}.

\item   $\mathcal{N}=2$ supergravity coupled to a single vector multiplet via a cubic pre-potential (often referred to as  the $T^{3}$ model and most simply obtained by  dimensionally  reducing  minimal $D=5$ supergravity):   the U-duality group is $%
\SU(1,1)$, with no additional   global
compact factors  present. The Abelian $2$-form field strengths and their duals sit in the $\mathbf{4}$ of $\SU(1,1)$. The manifold $%
\SU(1,1)/\Un(1)$ is an isolated case in the classification of symmetric special
K\"{a}hler spaces \cite{Cecotti:1988qn,deWit:1992wf}. Indeed, under dimensional reduction to
$D=3$, this is mapped to the exceptional quaternionic manifold $G_{2(2)}/\SO(4)$. The pair $\left(\SU(1,1),\mathbf{4}\right) $ provides the simplest
example of a group  of type $E_{7}$. \end{enumerate}
Although all share the same scalar coset none are twin. Firstly, the $\N=4$ theory has 32 degrees of freedom, while the two $\N=2$ theories each have 16. Secondly, despite having the same  bosonic (and fermionic) content and scalar manifolds the two $\N=2$ theories have distinct couplings in the bosonic sector since the Abelian field strengths  and their duals transform in two distinct representations of $\SU(1,1)$, the $\mathbf{2}+\mathbf{2}$ and $\mathbf{4}$, respectively. 

Similarly, for $\N=1$ supergravity the scalar coset $\SU(1,1)/\Un(1)$ considered above also appears in at least three examples. These belong to five families of $\N=1$ supergravity theories with scalar manifolds compatible with non-trivial electromagnetic duality  as given in \cite{Andrianopoli:2007rm}:
\begin{enumerate}
\item $\USp(2n, \R)/\Un(n)$ coupled to $n$ vector multiplets, with duality group $\Sp(n, \R)$.
\item $\Un(1,n)/\Un(n)$ coupled to $n + 1$ vector multiplets, with duality group $\Un(1, n) \subset \Sp(n + 1, \R)$.
\item $\SU(1,1)/\Un(1)$ coupled to $n$ vector multiplets, with duality group $\SL(2,\R)\times \SO(n) \subset \Sp(n, \R)$
\item $\SO(2, n)/\SO(2)\times \SO(n)$ coupled to $r$ vector multiplets in the $r$-dimensional spinor representation of $\SO(1, n-1)\subset\SO(2, n)$, with duality group $\Spin(2, n)\subset\Sp(r, \R)$
\item $\Un (n, n)/\Un(n)\times\Un(n)$ coupled to $2n$ vector multiplets.
\end{enumerate}
The scalar  coset $\SU(1,1)/\Un(1)$ occurs for: $(i)$ classes 1 and 4 coupled to a single vector multiplet with U-duality group $\Sp(1, \R)\cong\SU(1,1)$, $(ii)$ classes 2 and 5 coupled to two vector multiplets with U-duality group $\Un(1,1)\subset\Sp(4,\R)$ and $(iii)$ class 3 coupled to $n$ vector multiplets with U-duality group $\SL(2,\R)\times \SO(n)$. Note, case $(ii)$ has an additional $\Un(1)$ under which the scalars are neutral. The complex scalar of case $(iii)$ does not exhibit an attractor behaviour for spherically symmetric stationary black hole solutions \cite{Andrianopoli:2007rm}. 

These $\N=1$ theories with $\SU(1,1)/\Un(1)$ coset do constitute twins of  pure $\N=4$ supergravity and the $\N=2$ axion-dilaton model, but not the $T^3$ model as its kinetic vector matrix is not holomorphic, see for example (2.8) and (2.9) of \cite{Ceresole:2007rq}. In particular, case $(iii)$ with  $n=6$ is twinned with pure $\N=4$ supergravity. Dimensional reduction  to $D=3$, accompanied by the dualisation of all vectors, maps  the scalar  $D=4$ coset $\SU(1,1)/\Un(1)$ to  $\SO(8,2)/[\SO(8)\times \SO(2)]$. Similarly, case $(ii)$ is twinned with the  $\N=2$ axion-dilaton model, but in this case the coset $\SU(1,1)/\Un(1)$ is mapped to the quaternionic K\"{a}hler manifold $\SU(2,2)/[\SU(2)\times \SU(2)\times \Un(1)]$ in $D=3$.

\begin{table}[h!]
$\begin{array}{crcrrrcrcrrrcrc}
\toprule
&&&&~~\vline~~&&&&&&~~\vline~~&&&&\\
~~(\susy_+,\susy_-)_{(4)}~~ && \mathcal{M}_{scalar(4)} && ~~\vline~~ && ~~(\susy_+,\susy_-)_{(5)}~~
&& \mathcal{M}_{scalar(5)} && ~~\vline~~ && ~~(\susy_+,\susy_-)_{(6)}~~ &&
\mathcal{M}_{scalar(6)}\\
&&&&~~\vline~~&&&&&&~~\vline~~&&&&\\
\hline
\hline
&&&&~~\vline~~&&&&&&~~\vline~~&&&&\\
(2,1) && \frac{\Un(1,p-1)}{\Un(1)\times{\Un(p-1)}}\times\frac{\SU(2,q)}{\SU(2)\times{\Un(q)}}&&~~\vline~~&&&&&&~~\vline~~&&\\
&&&&~~\vline~~&&&&&&~~\vline~~&&&&\\
(3,1) && \frac{\Un(3,1)}{\Un(1)\times{\Un(3)}}&&~~\vline~~&&&&&&~~\vline~~&&\\
&&&&~~\vline~~&&&&&&~~\vline~~&&&&\\
(4,1) && \frac{\SU(1,1)}{\Un(1)}&&~~\vline~~&&&&&&~~\vline~~&&\\
&&&&~~\vline~~&&&&&&~~\vline~~&&&&\\
(5,1) && \frac{\SU(5,1)}{\Un(5)}&&~~\vline~~&&&&&&~~\vline~~&&\\
&&&&~~\vline~~&&&&&&~~\vline~~&&&&\\
(4,2) && \frac{\SU(1,1)}{\Un(1)}\times\frac{\SO(6,2)}{\Un(4)} &&~~\vline~~&&
(4,2) && \SO(1,1)\times\frac{{\SO(5,1)}}{\USp(2)} &&~~\vline~~&& ((1,1),(0,1))
&& \frac{\Orth(1,1)\times{\Sp(1)^2}}{\Un(1)^2}\\
&&&&~~\vline~~&&&&&&~~\vline~~&&&&\\
&&&&~~\vline~~&&&&&&~~\vline~~&& ((2,0),(0,1)) && \frac{\SU^\star(4)}{\USp(2)}\\
&&&&~~\vline~~&&&&&&~~\vline~~&&&&\\
(6,2) && \frac{\SO^\star(12)}{\Un(6)} &&~~\vline~~&& (6,2) && \frac{\SU^\star(6)}{\USp(3)}
&&~~\vline~~&& ((2,1),(0,1)) && \frac{\SU^\star(4)\times{\Sp(1)}}{\USp(2)\times{\Un(1)}}\\
&&&&~~\vline~~&&&&&&~~\vline~~&&&&\\
\hline
\hline
\end{array}$
\label{TWINS456}
\caption{The twin supergravities theories in $D=4,5,6$.}
\end{table}

The classification of twin supergravity theories is provided in  \autoref{TWINS456}. First we oxidise the twin pairs to $D=4$. At this point one would expect to get an infinite sequence of $(3,1)$ pairs with scalar manifold $\Un(3,p-1)/\Un(1)\times{\Un(3)}\times{\Un(p-1)}$. The $\susy=3$ theory is coupled to $p-1$ vector multiplets. The kinetic vector matrix of matter-coupled $\mathcal{N}=3$, $D=4$ supergravity has been computed in Appendix C of  \cite{Castellani:1985ka}, as well as more recently in \cite{Ferrara:2008ap} (\textit{cfr.} Section 4 and in particular (4.10) and (4.14) therein). As following from \cite{Castellani:1985ka, Ferrara:2008ap}  this matrix is always non-holomorphic, except in the case of one vector multiplet\footnote{We would like to thank Alessandra Gnecchi for useful correspondence concerning this point.}. Thus the only $(3,1)$ pair is the one belonging to the triplet $(3,2,1)$ with scalar coset $\Un(3,1)/\Un(3)\times{\Un(1)}\times{\Un(1)}$, again illustrating the point that for $D>3$ matching  bosonic content and scalar cosets is not sufficient.
Oxidising the twin pairs to $D=5$ is straightforward. Each twin pair oxidises to a unique pair of $D=5$ twin supergravities. Finally, we oxidise to $D=6$. One needs to be careful as both theories in the $(4,2)$ pair admit two different oxidations. The $\susy_-=2$ theory can be interpreted as part of the sequence $L(P,0)$ (with $P=4$)\footnote{This theory is anomaly-free only when coupled to 248 hyper multiplets.} or as part of the sequence $L(q,0)$ (with $q=4$)\footnote{This theory is anomaly-free
only when coupled to 128 hyper multiplets.} \cite{Gunaydin:1983bi, deWit:1984wbb, deWit:1992wf}. In the former case it uplifts to chiral $(0,1)$ theory coupled to $1$ tensors and $4$ vector multiplets, with U-duality group $\Orth(1,1)\times{\SO(4)}$. In the latter case it uplifts again to $(0,1)$ but now coupled to $5$ tensor multiplets and U-duality group $\SU^\star(4)$. The $\susy_+=4$ partner can uplift to either a $(2,0)$ or a $(1,1)$ theory \cite{Andrianopoli:1996ve}. The former is coupled to $1$ tensor multiplet while the latter is ``pure" supergravity. The respective U-duality groups are $\Orth(1,1)\times{\SO(4)}$ and $\SU^\star(4)$. This explains the two slots appearing in the last column of  \autoref{TWINS456}.

\section{Twins from super Yang-Mills squared}\label{twinsquare}

 In this section we describe the Yang-Mills squared origin of the twin supergravities. We begin by considering the pyramid of supergravity theories given in \autoref{PYR}. It is generated by the product of a left and right super Yang-Mills multiplet in $3\leq D\leq 10$,
\be\label{fact}
\bV_\nl\otimes \tbV_\nr = \bG_{\nl+\nr} \oplus \bM_{\nl+\nr},
\ee
where $\bV, \tbV$ are in the adjoint representation of $G$ and $\tG$. See \cite{Borsten:2013bp, Anastasiou:2013hba, Anastasiou:2015vba} for details. When a point in the pyramid $\bG_{\nl+\nr} \oplus \bM_{\nl+\nr}$ admits a twin we will denote the pair by 
\be
\bG_{\N_+} \oplus \bM_{\N_+}\quad  \underset{\text{twins}}{\longleftrightarrow} \quad \bG_{\N_-} \oplus \bM_{\N_-},
\ee
or simply
$
(\N_+, \N_-)$. In \autoref{PYR} we have summarised the twin pairs $(\N_+, \N_-)$ appearing in the pyramid. Note, since there are no twins for $D>6$ we have truncated the pyramid at $D=6$. We see that all theories obtained via  \eqref{fact},  excluding the maximal supergravities living on the spine for $D>3$, have a twin.

Not only do all the non-maximal supergravity theories obtained as the square of pure super Yang-Mills admit a twin, the two   theories are related in a controlled manner through their double-copy constructions. First note that $(\N_+, \N_-)$ can be regarded as a pair of complementary consistent truncations of a single $\mathscr{N}$-extended parent  supergravity theory  given by, 
\be
 \bV_\nl \otimes \tbV_\nr = \bG_{\nl+\nr}\oplus\bM_{\nl+\nr},
\qquad
\text{where} 
\qquad
\mathscr{N}=\nl+\nr=\N_++\N_-.
\ee 
This follows from simple symmetry requirements. The $\N_+$ R-symmetry is necessarily a subgroup of the parent $\mathscr{N}$ R-symmetry\footnote{When $\mathscr{N}\leq 4$ there can be an additional  isotropy group, which complicates the argument but does not change the conclusion.}. On the other hand, for the $\N_-$ twin the same group is  repurposed as matter isotropy group and we have to further include the $\N_-$ R-symmetry in the parent R-symmetry implying that the minimal $\mathscr{N}$ is given by $\N_++\N_-$.

Let us now summarise the twin double-copy procedure for  $(\N_+, \N_-)$:
\begin{enumerate} 
\item Decompose 
$
\bV_\nl=\bV_{\nl'}\oplus \bC_{\nl'}
$
where $\N_+=\nl'+\nr.$
\item Replace the adjoint spinor multiplet  by a fundamental   spinor multiplet,
$
\bC_{\nl'}\rightarrow \bC_{\nl'}^{\rho},
$
carrying a representation $\rho$ of the left gauge group and thus reducing the degree of supersymmetry to $\nl'<\nl$. 
\item  The double-copy construction
 \be
[\bV_{\nl'}\oplus \bC^{\rho}_{\nl'}]\otimes \tbV_{\nr}=\bV_{\nl'}\otimes \tbV_{\nr},
\ee
 yields the $\N_+=\nl'+\nr$ twin as a truncation of the parent supergravity through its Yang-Mills factors by discarding the states that would have arisen from $\bC_{\nl'}\otimes \tbV_{\nr}$, in particular  $\nl-\nl'$ of the gravitini.
\item Decompose 
\be\label{rdecomp}
\tbV_\nr=\tbV_{\nr'}\oplus \tbC_{\nr'}\oplus\cdots
\ee
where $\N_-=\nl'+\nr'$. Note that typically $\nr'=0$ so that $\tbV_\nr$ decomposes as
\be
\tbV_\nr=\tA\oplus\tX\oplus\tphi,
\ee
where we regard $\chi$ as an $\N=0$ spinor multiplet $\bC_0$.
\item Replace all adjoint spinor multiplets  by fundamental   spinor multiplets,
$
\tbC_{\nr'}\rightarrow \tbC_{\nr'}^{\tilde{\rho}} 
$ 
carrying a representation $\tilde{\rho}$ of the right gauge group $\tG$ and thus reducing the degree of supersymmetry to $\nr'<\nr$.
\item  The double-copy construction
 \be
[\bV_{\nl'}\oplus \bC^{\rho}_{\nl'}]\otimes [\tbV_{\nr'}\oplus \tbC^{\tilde{\rho}}_{\nr'}\oplus\cdots]=[\bV_{\nl'}\otimes \tbV_{\nr'}]\oplus[\bC^{\rho}_{\nl'}\otimes \tbC_{\nr'}^{\tilde{\rho}}]\oplus[\bV_{\nl'}\otimes \cdots],
\ee
 yields the $\N_-=\nl'+\nr'$ twin as a truncation of the parent supergravity through its Yang-Mills factors. Note, by using $\tbC_{\nr'}^{\tilde{\rho}}$ we discard $\nr-\nr'$ out of the $\nl'+\nr$ gravitini of the $\N_+$ twin that would have arisen from $\bV_{\nl'}\otimes\tbC_{\nr}$, as well as a subset of the spinor states. They are replaced by the spinors arising from $\bC^{\rho}_{\nl'}\otimes \tbC_{\nr'}^{\tilde{\rho}}$.
\end{enumerate} 
In summary   the two twin theories with $\N_+=\nl'+\nr$ and  $\N_-=\nl'+\nr'$ are related through,
\be\label{break1}
 [\bV_{\nl'}\oplus\bC_{\nl'}^{\rho}] \otimes \tbV_\nr \longleftrightarrow [\bV_{\nl'}\oplus\bC_{\nl'}^{\rho}] \otimes [\tbV_{\nr'}\oplus\bC_{\nr'}^{\tilde{\rho}}\oplus\cdots].  
\ee
The resulting twin theories in $D=3,4,5,6$  are given in \autoref{D3},  \autoref{D4},  \autoref{D5} and  \autoref{D6}.  The $D=4, (6,2)$ example is given in full detail in \autoref{example}, where some of the subtleties  of the above sketch  are addressed.    In particular:
\begin{itemize}
\item[$a)$] Multiplets carrying the adjoint (fundamental) representation of the left gauge group only double-copy with  multiplets carrying the adjoint (fundamental) representation of the right gauge group, leading to a \emph{sum of squares}. For example: 
\be\label{sum}
[\bV_\nl \oplus \bC^{\rho}_\nl]\otimes[\tbV_\nr \oplus \tbC^{\tilde{\rho}}_\nr] = [\bV_\nl\otimes \tbV_\nr]\oplus [ \bC^{\rho}_\nl\otimes\tbC^{\tilde{\rho}}_\nr].
\ee
This reflects the BCJ double-copy structure with fundamental matter \cite{Johansson:2014zca}.  As a consequence, although the degrees of freedom of the  Yang-Mills theory are not preserved, the twin supergravity theories  generated always have the same number of degrees of freedom as they must. 

\item[$b)$]  In verifying the double-copy relations one must account not only for the content of each theory, but also their symmetries. In particular, for supergravity coupled to matter multiplets one has to trace the Yang-Mills origin of both the  R-symmetry and the isotropy group of the matter  multiplets. This point is illustrated by the $D=4$ $\R, \C, \Q, \Oct$ magic supergravities. Their field content  is reproduced by the simple product,
\be
\bV_2\otimes[\tA \oplus  n \tphi].
\ee
 for $n= 5, 8, 14, 26$. However, the symmetries generated  are those of the generic Jordan sequence $\SL(2, \R)\times\SO(2, n)$  and not the magic sequence $\USp(3, \R), \SU(3,3), \SO^\star(12), E_{7(-25)}$ \cite{Anastasiou:2013hba}. The magic supergravities were double-copy constructed  in \cite{Chiodaroli:2015wal} using a right multiplet given by dimensionally reducing $\N=0$ Yang-Mills coupled to fundamental fermions in $D=7,8,10, 15$. 
 In the present construction the R-symmetry is always straight-forwardly generated by the left and right R-symmetries, but  the isotropy group  is more subtle. In particular it can place restrictions on the properties of the left and right gauge group representations, $\rho$ and $\tilde{\rho}$. For example, the $D=6, \N_-=2$ theory requires $\rho$ to be a real representations of $G$ so as to generate an enhanced flavour symmetry which in turn generates a part of the matter isotropy group. 
\end{itemize}

Before moving on to the $D=4, (6,2)$ example, let us briefly return to the ``sum of squares rule'' noted the above. Adjoint and fundamental representations do not mix in the double-copy prescription:
 \be\label{sum2}
 \begin{split}
[\bV_\nl \oplus \bC^{\rho}_\nl]\otimes[\tbV_\nr \oplus \tbC^{\tilde{\rho}}_\nr] &= [\bV_\nl\otimes \tbV_\nr]\oplus{\cancelto{\varnothing}{[\bV_\nl\otimes \tbC^{\tilde{\rho}}_{\nr}]}}\oplus{\cancelto{\varnothing}{[\bC^{\rho}_\nl\otimes \tbV_\nr]}}\oplus [ \bC^{\rho}_\nl\otimes\tbC^{\tilde{\rho}}_\nr]\\[5pt]
&= [\bV_\nl\otimes \tbV_\nr]\oplus [ \bC^{\rho}_\nl\otimes\tbC^{\tilde{\rho}}_\nr].
\end{split}
\ee
This is implied by supersymmetry as the cross-terms $\bV\otimes\tbC^{\tilde{\rho}}$ would introduce too many gravitini.  

 It also follows, with and without supersymmetry, from the structure of colour-kinematic duality for Yang-Mills coupled to fundamental\footnote{Note, the colour-kinematic duality for abelian orbifolds of   $D=4, \N=4$ super Yang-Mills theory with matter fields in a bi-fundamental representation was studied in \cite{Chiodaroli:2013upa}.} matter and hence the associated double-copy relations, as described in   \cite{Johansson:2014zca}.   For adjoint fields colour-kinematic duality is mediated by  the Jacobi identity. Of course, the Jacobi identities  are just the commutation relations in the adjoint representation, which immediately suggests the appropriate generalisation of colour-kinematic duality  to non-adjoint matter \cite{Johansson:2014zca}. Colour-kinematic duality for fundamental fields is mediated by the commutation  relations.   In summary, we have 
\be\label{BCJrels}
[f^{a}]_{c}{}^{d}[f^{b}]_{d}{}^{e}-[f^{b}]_{c}{}^{d}[f^{a}]_{d}{}^{e}=f^{ab}{}_{d}[f^{d}]_{c}{}^{e}\quad
\text{versus}\quad
[T^{a}]_{i}{}^{j}[T^{b}]_{j}{}^{k}-[T^{b}]_{i}{}^{j}[T^{a}]_{j}{}^{k}=if^{ab}{}_{d}[T^{d}]_{i}{}^{k},
\ee
where $[f^{a}]_{c}{}^{d}=f^{cad}$, $a=1,2, \ldots \dim G$ are the structure constants and $[T^{a}]_{i}{}^{j}$, $i,j=1, 2, \ldots \dim \rho(G)$ are the generators  in the appropriate matter-field representations. Since the colour-kinematic duality applies to triples of graphs with colour factors satisfying either the Jacobi or commutator identities, the fundamental matter  multiplets only  double-copy amongst themselves and, hence, the product of $\bC^{\rho}_\nl$ with $\tbV_\nr$ is trivial. 
 
To be more concrete consider an $n$-point, $L$-loop Yang-Mills-matter amplitude, which can be written in terms of trivalent graphs,
\be\label{amp}
A_{\1}{}_{n}^{L}=i^Lg^{n-2+2L}\sum_{i}\int \prod^{L}_{l=1}\frac{d^Dp_l}{(2\pi)^DS_i} \frac{ n_i c_i}{\prod_{a_i}p_{a_i}^{2}}.
\ee
The sum is over all $n$-point $L$-loop graphs $i$ with only trivalent vertices. For Yang-Mills coupled to fundamental matter  there are two possible classes of trivalent vertex: gluon-gluon-gluon ($a$,  $b$,  $c$)  dressed with a structure constant $f^{abc}$ or gluon-fund-antifund ($a$,  $i$,  $\bar{\jmath}$) dressed with a generator $[T^{a}]_{i}{}^{j}$. The colour factor of graph $i$  is denoted $c_i$. They are composed of gauge group structure constants $f^{abc}$ and generators $[T^{a}]_{i}{}^{j}$ and can be read-off the graph. The kinematic factor of graph $i$ is denoted $n_i$. It is a polynomial of Lorentz-invariant contractions  of polarisation vectors and momenta and includes any other quantum numbers, which must be specified. The $p_{a_i}^{2}$ are the propagators for each graph $i$. $S_i$ is the dimension of the automorphism group of graph $i$. The set of $n$-point trivalent graphs can always be organised into triples $i, j, k$ such that the colour factors will obey
\be
c_i+c_j+c_k=0,
\ee 
due to either the Jacobi  or commutation relations \eqref{BCJrels}.

For adjoint-valued fields (without fundamental matter) it was proposed in \cite{Bern:2008qj} that one can arrange the diagrams to display a remarkable colour-kinematic duality:
\be
\begin{array}{ccccc}
c_i+c_j+c_k=0&\Rightarrow& n_i+n_j+n_k=0
\end{array}
\ee
and if $c_i\rightarrow-c_i$ under the interchange of two legs then   $n_i\rightarrow-n_i$. A reorganisation admitting this surprising  relationship between  colour and kinematic data was shown to exist
  for all $n$-point tree-level amplitudes in \cite{Bern:2010yg}.  The colour-kinematic duality is conjectured to hold, with highly non-trivial evidence \cite{Bern:2009kd, Bern:2014sna}, at any loop level.  This colour-kinematic duality has been  extended to include fundamental matter multiplets using  commutation relations in place of the Jacobi identity  \cite{Johansson:2014zca}.

Assuming one has found a colour-kinematic duality respecting representation of the $n$-point $L$-loop (super) Yang-Mills-matter amplitude, mediated by both the Jacobi and commutation relations, the equivalent $n$-point $L$-loop (super)gravity amplitude $A_{\2}{}_{n}^{L}$ is obtained by simply replacing each colour factor, $c_i$, with a second kinematic factor, $\tilde{n}_i$   \cite{Bern:2008qj, Bern:2010yg, Bern:2010ue, Johansson:2014zca}:
\be\label{amp}
A_{\1}{}_{n}^{L}=i^Lg^{n-2+2L}\sum_{i}\int \prod^{L}_{l=1}\frac{d^Dp_l}{(2\pi)^DS_i} \frac{ n_i c_i}{\prod_{a_i}p_{a_i}^{2}}\quad\longrightarrow \quad i^L\left(\frac{\kappa}{2}\right)^{n-2+2L}\sum_{i}\int \prod^{L}_{l=1}\frac{d^Dp_l}{(2\pi)^DS_i} \frac{ n_i \tilde{n}_i}{\prod_{a_i}p_{a_i}^{2}}=A_{\2}{}_{n}^{L}.
\ee

For pure (super) Yang-Mills at tree-level it was shown  in \cite{Cachazo:2013iea} that one can also proceed in the other direction by replacing the kinematic factor  $n_i$ by a second colour factor $\tilde{c}_i$ to obtain a spin-0 amplitude $A_{\0}{}_{n}^{L}$ with two independent colour numerators. The result is the $n$-point tree-level amplitude of a massless bi-adjoint scalar field with cubic interaction,
\be
\mathcal{L}_{\text{bi-adj}}=-\frac{1}{2}\partial_\mu \Phi_{a\tilde{a}} \partial^\mu \Phi^{a\tilde{a}}+\frac{\lambda}{6}f_{abc}\tilde{f}_{\tilde{a}\tilde{b}\tilde{c}}\Phi^{a\tilde{a}} \Phi^{b\tilde{b}} \Phi^{c\tilde{c}},
\ee
where $f_{abc}$ and $\tilde{f}_{\tilde{a}\tilde{b}\tilde{c}}$ are the structure constants of two independent gauge groups $G$ and $\tG$. This has been  referred to as the zeroth-copy of Yang-Mills \cite{Monteiro:2014cda}. In this sense the $\Phi^3$ theory captures the colour structure of the left and right Yang-Mills factors as well as their propagators, which are common to the equivalent gravitational, gauge and scalar amplitudes.  

This   bi-adjoint  scalar field in fact plays  a ubiquitous role in various `gravity = gauge $\otimes$ gauge' constructions \cite{Hodges:2011wm, Cachazo:2013iea, Anastasiou:2014qba, Monteiro:2014cda}.   In the context of amplitudes its appearence is perhaps most clearly expressed in terms of the Cachazo-He-Yuan (CHY) formulae \cite{Cachazo:2013iea}.  Remarkably, the  spectator field plays a directly analogous role in the double-construction of classical black hole solutions \cite{Monteiro:2014cda, Luna:2015paa, Luna:2016due, White:2016jzc}. Finally, it was shown to be a crucial element of the linearised \emph{off-shell} dictionary, presented in \cite{Anastasiou:2014qba}, describing gravitational fields in terms of a convolutive tensor product of left and right Yang-Mills fields,
\be\label{offshell}
A_\mu^{a} \circ \Phi_{a\tilde{a}} \circ \tilde{A}_\nu^{\tilde{a}} = g_{\mu\nu} + B_{\mu\nu}.
\ee
Here $\circ$   denotes a convolutive inner tensor product with respect to the Poincar\'e group,
\be
[f\circ g](x)=\int d^Dy f(y) g(x-y).
\ee
In this context the ``spectator'' scalar field ensures that the symmetries of the (super) Yang-Mills factors are correctly mapped to those of (super)gravity. The convolution reflects the fact that the amplitude relations are multiplicative in momentum space. It turns out to be essential for reproducing the local symmetries of (super)gravity from those of the two (super) Yang-Mills factors to linear order. The spectator field allows for arbitrary and independent   $G$ and $\tG$ at the level of spacetime fields. Note, the bi-adjoint scalar field also appears by close analogy  in  the double-copy construction of classical black hole solutions  \cite{Monteiro:2014cda, Luna:2015paa, Luna:2016due, White:2016jzc}, although the precise relationship between the two pictures remains an intriguing open question.

\begin{figure}
\includegraphics[width=0.65\textwidth]{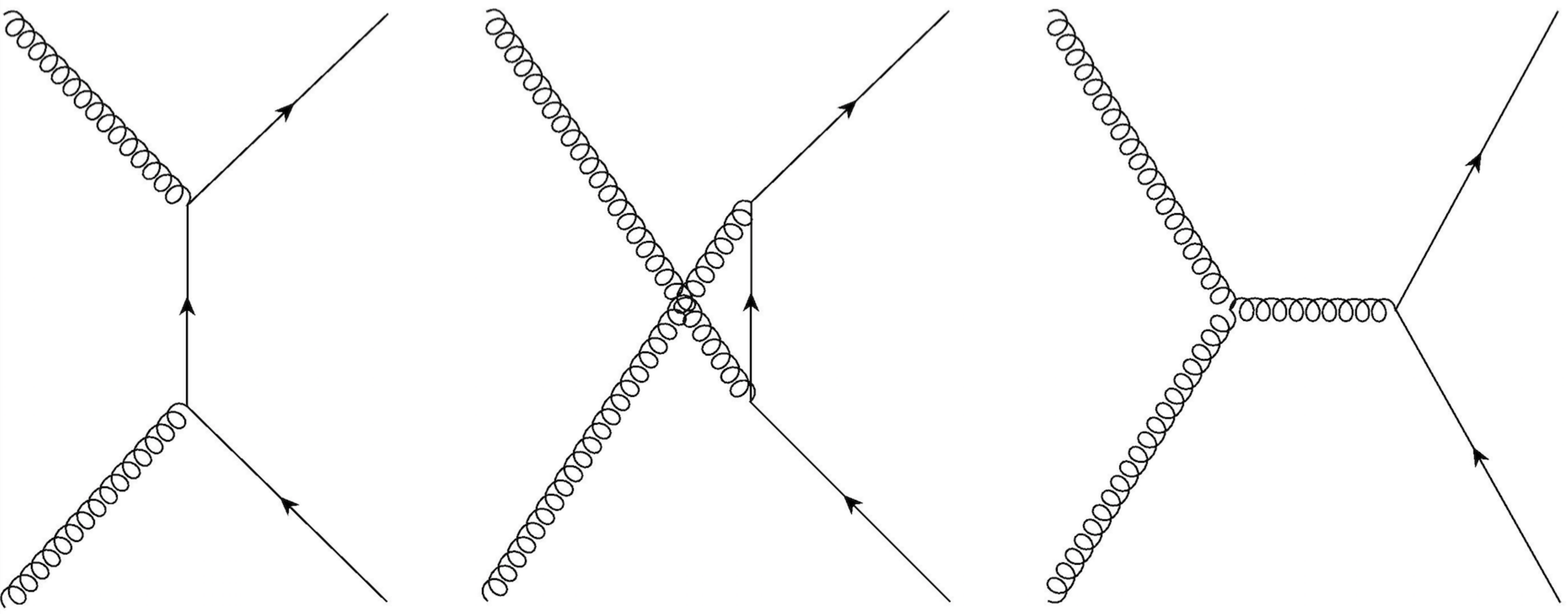}
\caption{Tree-level Feynman diagrams for the  4-point gluon-gluon-quark-antiquark interaction.}\label{feynmans}
\end{figure}

\begin{figure}
\includegraphics[width=0.65\textwidth]{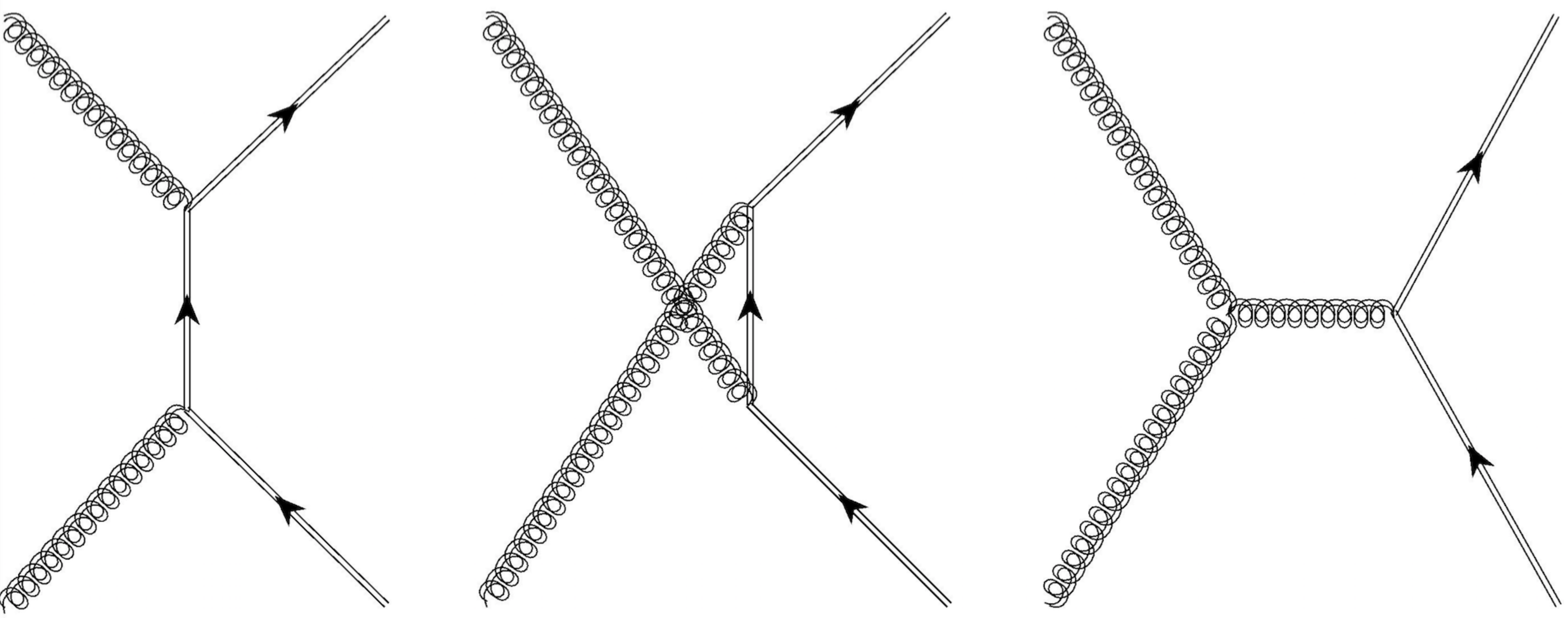}
\caption{Double-line tree-level Feynman diagrams for the  4-point (bi-adj.)-(bi-adj.)-(bi-fund.)-(bi-fund.)~interaction for the scalar theory in \eqref{bifundscalar}. The curly (straight) double-lines represent the bi-adjoint (bi-fundamental) representation of the global $G\times \tG$ symmetry. Each diagram is the double-copy of the corresponding gluon-quark diagram shawn of its kinematic data.}\label{feynmans1}
\end{figure}

Now, just as the generalised   colour-kinematic duality for adjoint and fundamental multiplets can be used to generate (super)gravity amplitudes with matter couplings, we can once again also proceed in the other direction (at tree-level at least) to generate  would-be spin-0 amplitudes. One can then look for the minimal scalar field theory that would produce these amplitudes. The result corresponds to the bi-adjoint scalar theory, but this time coupled to a  bi-fundamental scalar field $\Phi_{i\tilde{i}}$ with a cubic interaction term originating from the adj-fund-antifund vertices:
\be\label{bifundscalar}
\mathcal{L}_{\text{bi-adj-fund}}=-\frac{1}{2}\partial_\mu \Phi_{a\tilde{b}} \partial^\mu \Phi^{a\tilde{b}}-\frac{1}{2}\partial_\mu \Phi_{i\tilde{i}} \partial^\mu \Phi^{i\tilde{i}}+\frac{g}{6}\left(f_{abc}\tilde{f}_{\tilde{a}\tilde{b}\tilde{c}}\Phi^{a\tilde{a}} \Phi^{b\tilde{b}} \Phi^{c\tilde{c}}  +i[T^{a}]_{i}{}^{j}[{\tilde{T}}^{\tilde{a}}]_{\tilde{i}}{}^{\tilde{j}}\Phi_{a\tilde{a}} \Phi^{i\tilde{i}} \Phi_{j\tilde{j}}\right).
\ee
The defining example is given by the tree-level 4-point gluon-gluon-quark-antiquark interaction. The colour factors of the three Feynman diagrams, given in \autoref{feynmans}, obey the commutation relation \eqref{BCJrels}. 
As no gluon 4-point contact term is involved the BCJ representation of the kinematic numerators should, and does, follow directly from the Feynman rules. Using the labelling $(1, \varepsilon_{1}^{\mu}(k_1), a), (2, \varepsilon_{2}^{\nu}(k_2), b)$ for the two gluons  and $(3, v(k_3),  {}_i), (4, \bar{u}(k_4),  {}^j)$ for the quark-antiquark pair the three Feynman diagrams yield the colour-kinematic numerators:
\be
\begin{array}{llllllllll}
c_1 \times n_1&=&-i [T^{a}]_{i}{}^{k}[T^{b}]_{k}{}^{j}&\times&\bar{u}\varepsilon_{1}^{\mu}\gamma_\mu\gamma^\rho\left(k_4+k_1\right)_\rho\gamma_\nu\varepsilon_{2}^{\nu}v\\[5pt]
c_2 \times n_2&=&-i [T^{b}]_{i}{}^{k}[T^{a}]_{k}{}^{j}&\times&\bar{u}\varepsilon_{2}^{\nu}\gamma_\nu\gamma^\rho\left(k_4+k_2\right)_\rho\gamma_\mu\varepsilon_{1}^{\mu}v\\[5pt]
c_3 \times n_3&=&\phantom{-i} f^{abc}[T_{c}]_{i}{}^{j}&\times&\bar{u}\varepsilon_{1}^{\mu}\left(\eta_{\mu\rho}(k_1-p)_\nu+\eta_{\rho\nu}(p-k_2)_\mu+\eta_{\mu\nu}(k_2-k_1)_\rho\right)\varepsilon_{2}^{\nu}\gamma^\rho v
\end{array}
\ee
where we take all momenta $k_1, \ldots k_4$ to be out-going and $p=k_3+k_4$. Clearly $c_1-c_2=c_3$ and going on-shell we find $n_1-n_2=n_3$. The corresponding 4-point amplitude with two bi-adjoint and bi-fundamental scalar legs is then given by replacing $n_s$ with a copy  $\tilde{c}_s$, which need not carry the same gauge group, as given by double-line Feynman diagrams in \autoref{feynmans1}:
\be
\begin{array}{llllllllll}
c_1 \times \tilde{c}_1&=&-i [T^{a}]_{i}{}^{k}[T^{b}]_{k}{}^{j}&\times&[\tilde{T}^{\tilde{a}}]_{\tilde{i}}{}^{\tilde{k}}[\tilde{T}^{\tilde{b}}]_{\tilde{k}}{}^{\tilde{j}}\\[5pt]
c_2 \times \tilde{c}_2&=&-i [T^{b}]_{i}{}^{k}[T^{a}]_{k}{}^{j}&\times&[\tilde{T}^{\tilde{b}}]_{\tilde{i}}{}^{\tilde{k}}[\tilde{T}^{\tilde{a}}]_{\tilde{k}}{}^{\tilde{j}}\\[5pt]
c_3 \times \tilde{c}_3&=&\phantom{-i} f^{abc}[T_{c}]_{i}{}^{j}&\times&\tilde{f}^{\tilde{a}\tilde{b}\tilde{c}}[\tilde{T}_{\tilde{c}}]_{\tilde{i}}{}^{\tilde{j}}
\end{array}
\ee
 As for the double-copy, the denominators are  the same for both amplitudes.   Note, given a Feynman diagram representation, such as we have here, the   dictionary is quite intuitive. For each  vertex,  shorn of external states, the replacement rules are simply
\be
\label{repl1}
\gamma^\mu{}_\alpha{}^{\beta}\rightarrow [\tilde{T}^{\tilde{a}}]_{\tilde{i}}{}^{\tilde{j}} 
\ee
and
\be
\label{repl2}
\eta_{\mu_1\mu_2}(k_1-k_2)_{\mu_3}+\eta_{\mu_2\mu_3}(k_2-k_3)_{\mu_1}+\eta_{\mu_3\mu_1}(k_3-k_1)_{\mu_2}\rightarrow \tilde{f}^{\tilde{a}_1\tilde{a}_2\tilde{a}_3}.
\ee 
All contractions of gauge group indices amongst vertices are dictated by the contractions of the corresponding kinematic indices  by  propagators.
Note, we can accommodate a broader class of double-copy constructions by also including a fundamental-antifundamental scalar $\Phi^{i}{}_{\tilde{i}}$, allowing for distinct colour structures in the two factors mirroring the product of distinct kinematic structures in the  double-copy for gravitational amplitudes. This will be developed in future work. The scalar field theory can also be further generalised to accommodate structural features such as quark flavours. A 5-point example, following \cite{Johansson:2014zca},  illustrating this point is given in \aref{App5point}.

Most importantly for the present work is that the bi-fundamental scalar is consistent with spacetime field dictionary of  \cite{Anastasiou:2014qba}. In particular, we can generalise \eqref{offshell} to include fundamental fields in the left and right multiplets by introducing a block-diagonal spectator field $\Phi$ with bi-adjoint and bi-fundamental  sectors,
\be
\Phi = \begin{pmatrix} \Phi^{a\tilde{b}} & 0\\ 0& \Phi^{i\tilde{j}}\end{pmatrix}.
\ee
The Lorentz covariant position-space dictionary then correctly captures the sum-of-squares rule: 
\be
 [\bV_\nl\oplus \bC^{\rho}_\nl] \star  [\tbV_\nr\oplus \tbC^{\tilde{\rho}}_\nr] = [\bV_\nl\oplus \bC^{\rho}_\nl] \circ \Phi  \circ  [\tbV_\nr\oplus \tbC^{\tilde{\rho}}_\nr] =\bV_\nl^{a} \circ \Phi_{a\tilde{a}}  \circ \tbV_\nr^{\tilde{a}} \oplus \bC^{i}_\nl \circ \Phi_{i\tilde{i}} \circ  \tbC^{\tilde{i}}_\nr.
\ee
Crucially, the   symmetries of  the Yang-Mills-matter factors are correctly  mapped to the global and local (super)gravity symmetries via $\star$ and $\Phi$. This allows us to establish the corresponding supergravity theory through the  structure of the Yang-Mills symmetries alone, assuming that the gravitational scalar fields parametrise symmetric spaces. 

\subsection{Example: the $D=4, (6,2)$  twin theories}\label{example}

To make this procedure concrete let us consider in detail the prototypical example: $D=4, \N_+=6$ pure supergravity and its twin, the magic  $D=4, \N_-=2$ supergravity coupled to 15 vector multiplets.

The $D=4, \N=6$ supergravity theory is unique and determined by supersymmetry. The  multiplet consists  of
\be
\bG_6 = \{g_{\mu\nu}, 16 A_\mu, 30 \phi ;    6 \Psi_\mu, 26 \chi \}.
\ee
The R-symmetry algebra is $\mathfrak{u}(6)$ under which the on-shell helicity states transform as:
\be\label{n6}
\begin{array}{rrccccccccccccc}
&&\mathfrak{so}(2)&\mathfrak{u}(6)\\[4pt]
 & |& 2  &\rep{1}_0&  \rangle\\[4pt]
 Q & |& \frac{3}{2}  &\rep{6}_1&  \rangle\\[4pt]
 \wedge^2 Q & |& 1  &\rep{15}_2&  \rangle\\[4pt]
 \wedge^3 Q& |& \frac{1}{2}  &\rep{20}_3&  \rangle\\[4pt]
 \wedge^4 Q & |& 0  &\overline{\rep{15}}_4&  \rangle\\[4pt]
 \wedge^5 Q & |& -\frac{1}{2} &\overline{\rep{6}}_5&  \rangle\\[4pt]
 \wedge^6 Q & |& -1  &{\rep{1}}_6&  \rangle\\[4pt]
\end{array}\hspace{0.5in}
\begin{array}{rrccccccccccccc}
&&\mathfrak{so}(2)&\mathfrak{u}(6)\\[4pt]
 & |& 1  &\rep{1}_{-6}&  \rangle\\[4pt]
 Q & |& \frac{1}{2}  &\rep{6}_{-5}&  \rangle\\[4pt]
 \wedge^2 Q & |& 0  &\rep{15}_{-4}&  \rangle\\[4pt]
 \wedge^3 Q& |& -\frac{1}{2}  &\rep{20}_{-3}&  \rangle\\[4pt]
 \wedge^4 Q & |& -1  &\overline{\rep{15}}_{-2}&  \rangle\\[4pt]
 \wedge^5 Q& |& -\frac{3}{2} &\overline{\rep{6}}_{-1}&  \rangle\\[4pt]
 \wedge^6 Q & |& -2  &{\rep{1}}_{0}&  \rangle\\[4pt]
\end{array}
\ee
The non-compact global symmetry of the equations of motion is $\SO^\star(12)$.  The 15+15 scalars parametrise the coset manifold
$\SO^\star(12)/\Un(6)$,
\be
\begin{array}{lll}
\mathfrak{so}^\star(12)&\supset& \mathfrak{u}(1)\oplus\mathfrak{su}(6);\\
\rep{66}&\rightarrow& [\rep{1}+\rep{35}]_0 + \rep{15}_{-4} + \overline{\rep{15}}_{4}.
\end{array}
\ee
The 16 Maxwell field strengths and their duals transform as the $\rep{32}$ of $\SO^\star(12)$,
\be
\begin{array}{lll}
\mathfrak{so}^\star(12)&\supset& \mathfrak{u}(1)\oplus\mathfrak{su}(6);\\
\rep{32}&\rightarrow& \rep{1}_6+\rep{1}_{-6} + \rep{15}_{2} + \overline{\rep{15}}_{-2}.
\end{array}
\ee

The parent theory is $\N_++\N_-=8$ supergravity  with on-shell helicity states,
 \be
\begin{array}{rrccccccccccccc}
&&\mathfrak{so}(2)&\mathfrak{su}(8)\\[4pt]
 & |& 2  &\rep{1}&  \rangle\\[4pt]
 Q & |& \frac{3}{2}  &\rep{8}&  \rangle\\[4pt]
 \wedge^2 Q & |& 1  &\rep{28}&  \rangle\\[4pt]
 \wedge^3 Q & |& \frac{1}{2}  &\rep{56}&  \rangle\\[4pt]
 \wedge^4 Q & |& 0  &\rep{70}&  \rangle\\[4pt]
 \wedge^5 Q & |& -\frac{1}{2} &\overline{\rep{56}}&  \rangle\\[4pt]
 \wedge^6 Q & |& -1  &\overline{\rep{28}}&  \rangle\\[4pt]
 \wedge^7 Q & |& -\frac{3}{2}  &\overline{\rep{8}}&  \rangle\\[4pt]
 \wedge^8 Q & |& -2  &\rep{1}&  \rangle\\[4pt]
\end{array}
\ee
As a truncation the $\N_+=6$ theory  is obtained by decomposing with respect to $\mathfrak{su}(8)\supset\mathfrak{su}(6)\times\mathfrak{su}(2)\times\mathfrak{u}(1)$ and  discarding the non-trivial $\mathfrak{su}(2)$ representations,
\be
\begin{array}{llllll}
\rep{8}&\rightarrow&(\rep{1,2})_\text{-3}+(\rep{6,1})_\text{1}\\[5pt]
\rep{28}&\rightarrow&(\rep{1,1})_\text{-6}+(\rep{6,2})_\text{-2}+(\rep{15,1})_\text{2}\\[5pt]
\rep{56}&\rightarrow&(\rep{6,1})_\text{-5}+(\rep{15,2})_\text{-1}+(\rep{20,1})_\text{3}\\[5pt]
\rep{70}&\rightarrow&(\rep{15,1})_\text{-4}+\left(\rep{\overline{15},1}\right)_\text{4}+(\rep{20,2})_\text{0}\\[5pt]
\rep{\overline{56}}&\rightarrow&\left(\rep{\bar{6},1}\right)_\text{5}+\left(\rep{\overline{15},2}\right)_\text{1}+(\rep{20,1})_\text{-3}\\[5pt]
\rep{\overline{28}}&\rightarrow&(\rep{1,1})_\text{6}+\left(\rep{\bar{6},2}\right)_\text{2}+\left(\rep{\overline{15},1}\right)_\text{-2}\\[5pt]
\rep{\overline{8}}&\rightarrow&(\rep{1,2})_\text{3}+\left(\rep{\bar{6},1}\right)_\text{-1}\\[5pt]
\end{array}
\ee	

From the perspective of Yang-Mills squared we have the unique product,
\be
\bV_4\otimes\tbV_4=\bG_8 
\ee
and the $\N_+=6$ multiplet is the product of  $\N=2$ and $\N=4$ vector multiplets. The above truncation from $\N=8$ is effected by decomposing one of the $\N=4$ vector multiplets  (we choose the left) into an $\N=2$ vector-multiplet plus hyper-multiplet, 
\be\label{v4tov2h2}
\begin{array}{rrccccccccccccc}
&&\mathfrak{so}(2)&\mathfrak{su}(4)\\[4pt]
  & |& 1  &\rep{1}&   \rangle\\[4pt]
  Q& |& \frac{1}{2}  &\rep{4} &  \rangle\\[4pt]
 \wedge^2 Q & |& 0  &{\rep{6}} &  \rangle\\[4pt]
 \wedge^3 Q & |& -\frac{1}{2} &\overline{\rep{4}} & \rangle\\[4pt]
 \wedge^4 Q & |& -1  &{\rep{1}} & \rangle\\[4pt]
\end{array}\hspace{0.25in}\longrightarrow\hspace{0.25in}
\begin{array}{rrccccccccccccc}
&&\mathfrak{so}(2)&\mathfrak{u}(2)\oplus\mathfrak{su}(2)\\[4pt]
 & |& 1  &(\rep{1}, \rep{1})_0 &\rangle\\[4pt]
 Q& |& \frac{1}{2}  &(\rep{2},  \rep{1})_1& \rangle\\[4pt]
 \wedge^2 Q & |& 0  &({\rep{1}}, \rep{1})_2& \rangle\\[4pt]
 &&&\\
 & |& 0  &({\rep{1}}, {\rep{1}})_{-2}&  \rangle\\[4pt]
 Q & |& -\frac{1}{2} &({\rep{2}}, {\rep{1}})_{-1}&  \rangle\\[4pt]
 \wedge^2 Q & |& -1  &({\rep{1}}, {\rep{1}})_{ 0}&  \rangle\\[4pt]
\end{array}\hspace{0.1in}\oplus\hspace{0.1in}
\begin{array}{rrccccccccccccc}
&&\mathfrak{so}(2)&\mathfrak{u}(2)\oplus\mathfrak{su}(2)\\[4pt]
 & |& \frac{1}{2}   &(\rep{1},  \rep{2})_{-1} &\rangle\\[4pt]
 Q& |& 0 &(\rep{2},  \rep{2})_{0}& \rangle\\[4pt]
 \wedge^2 Q & |& -\frac{1}{2}   &({\rep{1}},  \rep{2})_{1}& \rangle\\[4pt]
\end{array}
\ee
where $\mathfrak{su}(4)_R\supset \mathfrak{u}(2)_R\oplus\mathfrak{su}(2)\oplus\mathfrak{u}(1)$. Rather than truncate $\mathbf{H}_2$ we  replace it with another hyper-multiplet in a fundamental representation $\rho$ of the gauge group,
\be
\bV_4\otimes \tbV_4 = \bG_8  \longrightarrow \left[\bV_2 \oplus \mathbf{H}_{2}^{\rho}\right]\otimes \tbV_4 = \bG_6,
\ee
 as indicated by the superscript. This reduces the left supersymmetry from $\nl=4$ to $\nl'=2$. To preserve the $\mathfrak{su}(2)\oplus\mathfrak{su}(2)$ symmetry  $\rho$ must be a real representation. The second $\mathfrak{su}(2)$ factor is an enhanced flavour symmetry that is only present for real representations of the gauge group. This is a special case of the enhanced $\mathfrak{sp}(n)$ or $\mathfrak{so}(n)$ flavour symmetry enjoyed by $n$ hypermultiplets in a real or pseudo-real gauge group representation, respectively. As $\mathbf{H}_{2}^{\rho}$ does not `talk' to the right adjoint valued multiplet $\tbV_4$, from the perspective of squaring it is effectively truncated. Since $\bV_2$ is a singlet under the $\mathfrak{su}(2)$ flavour it plays no role here either; this reflects the fact that   $\N=6$ supergravity does not admit matter couplings and there is no corresponding isotropy group.  The $\mathfrak{u}(6)$ of R-symmetry is, roughly speaking, generated by the left and right $\mathfrak{u}(2)$ and   $\mathfrak{su}(4)$ R-symmetries. Explicitly, we have 
\be\label{2x4ii}
\begin{array}{l|lllllllllllllllllllllllll}
\N'=2\backslash\tilde{\N}=4 && |1; \rep{1} \rangle && |\frac{1}{2}; \rep{4} \rangle &&|0; \rep{6} \rangle  && |\bar{\frac{1}{2}}; \overline{\rep{4}} \rangle&&  |\bar{1}; \rep{1} \rangle \\
&&&&&&&&&\\
\hline
&&&&&&&&&\\
|1; (\rep{1}, \rep{1})_0 \rangle && |2; (\rep{1}, \rep{1}, \rep{1})_{00} \rangle && |\frac{3}{2}; (\rep{1}, \rep{1}, \rep{4})_{0\frac{1}{2}} \rangle && |1; (\rep{1}, \rep{1}, \rep{6})_{01} \rangle && |\frac{1}{2}; (\rep{1}, \rep{1}, \overline{\rep{4}})_{0\frac{3}{2}} \rangle && |0; (\rep{1}, \rep{1}, \rep{1})_{02} \rangle \\[8pt]
  |\frac{1}{2}; (\rep{2}, \rep{1})_{1} \rangle &&  |\frac{3}{2}; (\rep{2},  \rep{1}, \rep{1})_{1\bar{\frac{1}{2}}} \rangle && |1; (\rep{2}, \rep{1}, \rep{4})_{10} \rangle && |\frac{1}{2}; (\rep{2}, \rep{1}, \rep{6})_{1\frac{1}{2}} \rangle && |0; (\rep{2}, \rep{1}, \overline{\rep{4}})_{11} \rangle && |\bar{\frac{1}{2}}; (\rep{2}, \rep{1}, \rep{1})_{1\frac{3}{2}} \rangle \\[8pt]
  |\frac{1}{2}; (\rep{1}, \rep{2})_{\bar{1}} \rangle_\rho &&   \\[8pt]

 |0; (\rep{1}, \rep{1})_2 \rangle   &&  |1; (\rep{1}, \rep{1},  \rep{1})_{2\bar{1}} \rangle && |\frac{1}{2}; (\rep{1},  \rep{1},  \rep{4})_{2\bar{\frac{1}{2}}} \rangle && |0; (\rep{1},  \rep{1},  \rep{6})_{20} \rangle && |\bar{\frac{1}{2}}; (\rep{1},  \rep{1},  \overline{\rep{4}})_{2\frac{1}{2}} \rangle && |\bar{1}; (\rep{1}, \rep{1},   \rep{1})_{21} \rangle \\[8pt]
  |0; (\rep{2}, \rep{2})_{0} \rangle_\rho &&   \\[8pt]
  |0; (\rep{1}, \rep{1})_{\bar{2}} \rangle  &&  |1; (\rep{1},  \rep{1},   \rep{1})_{\bar{2}\bar{1}} \rangle && |\frac{1}{2}; (\rep{1},  \rep{1},  \rep{4})_{\bar{2}\bar{\frac{1}{2}}} \rangle && |0; (\rep{1},  \rep{1},  \rep{6})_{\bar{2}0} \rangle && |\bar{\frac{1}{2}}; (\rep{1},  \rep{1},  \overline{\rep{4}})_{\bar{2}\frac{1}{2}} \rangle && |\bar{1}; (\rep{1},  \rep{1},  \rep{1})_{\bar{2}} \rangle \\[8pt]
    |\frac{1}{2}; (\rep{1},  \rep{1},  \rep{2})_{{1}} \rangle_\rho &&   \\[8pt]
  |\bar{\frac{1}{2}}; (\rep{2}, \rep{1})_{\bar{1}} \rangle  &&  |\frac{1}{2}; (\rep{2},   \rep{1},  \rep{1})_{\bar{1}\bar{\frac{3}{2}}} \rangle && |0; (\rep{2},  \rep{1},  \rep{4})_{\bar{1}\bar{1}} \rangle && |\bar{\frac{1}{2}}; (\rep{2},  \rep{1},  \rep{6})_{\bar{1}\bar{\frac{1}{2}}} \rangle && |\bar{1}; (\rep{2},  \rep{1},  \overline{\rep{4}})_{\bar{1}0} \rangle && |\bar{\frac{3}{2}}; (\rep{2},  \rep{1},  \rep{1})_{\bar{1}{\frac{1}{2}}} \rangle\\[8pt]
  |\bar{1}; (\rep{1}, \rep{1})_0 \rangle  &&  |0; (\rep{1},   \rep{1},  \rep{1})_{0\bar{2}} \rangle && |\bar{\frac{1}{2}}; (\rep{1},  \rep{1},  \rep{4})_{0\bar{\frac{3}{2}}} \rangle && |\bar{1}; (\rep{1},  \rep{1},  \rep{6})_{0\bar{1}} \rangle && |\bar{\frac{3}{2}}; (\rep{1}, \rep{1},   \overline{\rep{4}})_{0\bar{\frac{1}{2}}} \rangle && |\bar{2}; (\rep{1},  \rep{1},  \rep{1})_{00} \rangle \\
\end{array}
\ee
where for notational clarity we have used  $\bar{\lambda}\equiv-\lambda$ to denote negative helicities. In the above we have included the effectively trivial  states belonging to the fundamental hypermultiplet (as indicated by the $\rho$ subscript) to illustrate how the truncation of the $\mathscr{N}=8$ theory is effected.  Here the left $\nl=2$ super Yang-Mills states carry $\mathfrak{so}(2)_l \oplus\mathfrak{u}(2)_L\oplus\mathfrak{su}(2)_L$ spacetime little group, R-symmetry and flavour representations and the right $\N_R=4$ states carry $\mathfrak{so}(2)_r\oplus\mathfrak{su}(4)_R$ spacetime little group and R-symmetry representations. The $\N=6$ supergravity states carry $\mathfrak{so}(2)_{st}\oplus [\mathfrak{u}(2)_L\oplus\mathfrak{su}(2)_L] \oplus [\mathfrak{su}(4)_R]\oplus \mathfrak{u}(1)$ representations, where the spacetime helicity group $\mathfrak{so}(2)_{st}$ and the additional $\mathfrak{u}(1)$ factor are given by the sum and difference of the $\mathfrak{so}(2)_l$ and $\mathfrak{so}(2)_r$ generators, respectively. The charges carried by the extra $\mathfrak{u}(1)$ are given by the second subscript. Following \cite{Anastasiou:2015vba} the $[\mathfrak{u}(2)_L] \oplus [\mathfrak{su}(4)_R]\oplus \mathfrak{u}(1)$  generators are completed to the $\N=6$ R-symmetry algebras $\mathfrak{u}(6)$. All states are trivial under $\mathfrak{su}(2)_L$, which drops out of the equations. 

 Before the states can be assembled into the corresponding irreducible $\N=6$ multiplet we have to take a linear combination of the $\mathfrak{u}(1)_L$ and $\mathfrak{u}(1)$ generators,
\be
\begin{pmatrix} h_1 \\ h_2 \end{pmatrix} =\begin{pmatrix} 2 & -1 \\ 2& 2 \end{pmatrix}  \begin{pmatrix} h_L \\ h \end{pmatrix}. 
\ee
We then reproduce the states as given in \eqref{n6}, as can be seen by comparing \eqref{2x4ii}  with the  following decompositions:
\be
\begin{array}{llllll}
\mathfrak{su}(6)&\supset& \mathfrak{su}(2)\oplus\mathfrak{su}(4)\oplus\mathfrak{u}(1) \\
&&\\
\rep{6}&\rightarrow&\rep{(2,1)}_{-2}+\rep{(1,4)}_{1}\\[5pt]
\rep{15}&\rightarrow&\rep{(1,1)}_{-4}+\rep{(2, 4)}_{-1}+\rep{(1, 6)}_{2}\\[5pt]
\rep{20}&\rightarrow&\rep{(1, 4)}_{-3}+\rep{\left(1, \bar{4}\right)}_{3}+\rep{(2, 6)}_{0}\\[5pt]
\rep{\overline{15}}&\rightarrow& \rep{(1,1)}_{4}+\rep{\left(2, \bar{4}\right)}_{1}+\rep{(1, 6)}_{-2}\\[5pt]
\rep{\overline{6}}&\rightarrow&\rep{(2, 1)}_{2}+\rep{\left(1, \bar{4},\right)}_{-1}\\[5pt]
\end{array}
\ee

Its  little twin theory is the magic $\N=2$ supergravity coupled to 15 vector multiplets based on the Jordan algebra of $3\times 3$ Hermitian quaternionic matrices $\J_3(\Q)$, with content:
\be
\bG_2\oplus15\bV_2 = \{g_{\mu\nu},  A_\mu; 2 \Psi_\mu\}\oplus 15\{ A_\mu,  2\phi ;    2 \chi \}.
\ee
The R-symmetry algebra is $\mathfrak{u}(2)_R$ and the isotropy algebra is $\mathfrak{su}(6)$ under which the on-shell helicity states transform as:
\be\label{n2}
\begin{array}{rrccccccccccccc}
&&\mathfrak{so}(2)&\mathfrak{u}(2)_R\oplus\mathfrak{su}(6)\\[4pt]
 & |& 2  &(\rep{1}, \rep{1})_{0}&  \rangle\\[4pt]
 Q & |& \frac{3}{2}  &(\rep{2}, \rep{1})_{1}&  \rangle\\[4pt]
 \wedge^2 Q & |& 1  &(\rep{1}, \rep{1})_{2}&  \rangle\\[4pt]
 \\
  & |& {-1}  &(\rep{1}, \rep{1})_{-2}&  \rangle\\[4pt]
 Q & |& -\frac{3}{2}  &(\rep{2}, \rep{1})_{{-1}}&  \rangle\\[4pt]
 \wedge^2 Q & |& -2  &(\rep{1}, \rep{1})_{0}&  \rangle\\[4pt]
\end{array}\hspace{0.25in}\oplus\hspace{0.25in}
\begin{array}{rrccccccccccccc}
&&\mathfrak{so}(2)&\mathfrak{u}(2)_R\oplus\mathfrak{su}(6)\\[4pt]
 & |& 1  &(\rep{1}, \overline{\rep{15}})_{0}&  \rangle\\[4pt]
 Q & |& \frac{1}{2}  &(\rep{2}, \overline{\rep{15}})_{1}&  \rangle\\[4pt]
 \wedge^2 Q & |& 0  &(\rep{1}, \overline{\rep{15}})_{2}&  \rangle\\[4pt]
 \\
  & |& 0  &(\rep{1}, \rep{15})_{-2}&  \rangle\\[4pt]
 Q & |& -\frac{1}{2}  &(\rep{2}, \rep{15})_{-1}&  \rangle\\[4pt]
 \wedge^2 Q & |& -1  &(\rep{1}, \rep{15})_{0}&  \rangle\\[4pt]
\end{array}
\ee
  The 30 scalars parametrise the coset manifold
\be
\frac{\SU(2)}{\SU(2)}\times\frac{\SO^\star(12)}{\Un(6)}\cong\frac{\SO^\star(12)}{\Un(6)},
\ee
where the $\Un(1)$ of the R-symmetry has been ``gauged'' such that 
\be
\begin{array}{lll}
\mathfrak{so}^\star(12)&\supset& \mathfrak{u}(1)\oplus\mathfrak{su}(6);\\
\rep{66}&\rightarrow& [\rep{1}+\rep{35}]_0 + \rep{15}_{-4} + \overline{\rep{15}}_{4}.
\end{array}
\ee
The non-compact global symmetry of the equations of motion is $\SO^\star(12)$, under which the 16 Maxwell field strengths and their duals comprise the $\rep{32}$ spinor representation,
\be
\begin{array}{lll}
\mathfrak{so}^\star(12)&\supset& \mathfrak{u}(1)\oplus\mathfrak{su}(6);\\
\rep{32}&\rightarrow& \rep{1}_6+\rep{1}_{-6} + \rep{15}_{2} + \overline{\rep{15}}_{-2}.
\end{array}
\ee
As described in the original treatment of the magic supergravities \cite{Gunaydin:1983rk, Gunaydin:1983bi, Gunaydin:1984ak}, the   15 potentials and their duals can be regarded as elements of the Jordan algebra of $3\times3$ Hermitian matrices defined over the quaternions, $\J_3(\Q)$, and its dual with respect to the bilinear Jordan trace form,  $\J_3(\Q)^*\cong \J_3(\Q)$. Combined with the graviphoton and its dual these can be assembled into a 32-dimensional Freudenthal triple system and the pair $(\SO^\star(12), \rep{32})$ constitutes a group of type $E_7$.

To generate the magic $\N_-=2$  theory we similarly decompose the right $\nr=4$ multiplet into $\nr'=0$ multiplets,
\be
\bV_4=\{\tA, \tphi_{[\alpha\beta]}; \tX_\alpha^{\tilde{\rho}}\}
\ee
where $\alpha, \beta=1,\ldots4$ are indices of the fundamental of the R-symmetry remnant $\mathfrak{su}(4)$. Here we have  replaced the \emph{adjoint-valued}  $\tX_\alpha$ by a \emph{fundamental-valued} spinor $\tX_\alpha^{\tilde{\rho}}$ reducing the degree of supersymmetry. We then have the twin truncations of the parent $\mathscr{N}=8$ theory:
\be
    \xymatrix{
             & {\begin{array}{c}\mathscr{N}=8 ~\text{Parent supergravity}\\ \bG_{8}\end{array}}  \ar[d]^{\text{Yang-Mills factors}} &   \\
         & \bV_4 \otimes \tbV_4 \ar[dl] \ar[dr]  &   \\
         [\bV_{2}\oplus\bH_{2}^{\rho}] \otimes \tbV_4   \ar[d]   \ar@{<-->}[rr]_{\text{twin relation}}               &  &  [\bV_{2}\oplus\bH_{2}^{\rho}] \otimes [\tA\oplus\tX_{\alpha}^{\tilde{\rho}}\oplus\tphi_{[\alpha\beta]}]  \ar[d] \\
          \bV_{2} \otimes \tbV_4   \ar[d]                 &  &  [\bV_{2}\otimes (\tA\oplus\tphi_{[\alpha\beta]})]\oplus[\bH_{2}^{\rho}\otimes \tX_{\alpha}^{\tilde{\rho}}]  \ar[d] \\{\begin{array}{c}\N_+ =6 ~ \text{supergravity}\\ 
     \bG_{6}  \end{array}}    &  &{\begin{array}{c}\N_-=2 ~ \text{magic supergravity}\\  \bG_{2}\oplus\bV_{2}\oplus2\bV_{2\alpha}\oplus\bV_{2[\alpha\beta]}\end{array}}  }
\ee

Explicitly, we have the complementary truncation, cf. \eqref{2x4ii}, of $\mathscr{N}=8$ supergravity in terms of the left and right helicity states:
\be\label{2x0}
\begin{array}{l|lllllllllllllllllllllllll}
 \N'=2\backslash\tilde{\N}'=0&& |1; \rep{1} \rangle && |\frac{1}{2}; \rep{4} \rangle_{\tilde{\rho}} &&|0; \rep{6} \rangle  && |\bar{\frac{1}{2}}; \overline{\rep{4}} \rangle_{\tilde{\rho}}&&  |\bar{1}; \rep{1} \rangle \\
&&&&&&&&&\\
\hline
&&&&&&&&&\\
|1; (\rep{1}, \rep{1})_0 \rangle && |2; (\rep{1}, \rep{1},  \rep{1})_{00} \rangle && && |1; (\rep{1}, \rep{1}, \rep{6})_{01} \rangle && && |0; (\rep{1}, \rep{1}, \rep{1})_{02} \rangle \\[8pt]
  |\frac{1}{2}; (\rep{2}, \rep{1})_1 \rangle &&  |\frac{3}{2}; (\rep{2}, \rep{1}, \rep{1})_{1\bar{\frac{1}{2}}} \rangle && && |\frac{1}{2}; (\rep{2}, \rep{1}, \rep{6})_{1\frac{1}{2}} \rangle &&&& |\bar{\frac{1}{2}}; (\rep{2}, \rep{1}, \rep{1})_{1\frac{3}{2}} \rangle \\[8pt]
 |0; (\rep{1}, \rep{1})_2 \rangle   &&  |1; (\rep{1}, \rep{1}, \rep{1})_{2\bar{1}} \rangle &&  && |0; (\rep{1}, \rep{1}, \rep{6})_{20} \rangle && && |\bar{1}; (\rep{1}, \rep{1}, \rep{1})_{21} \rangle \\[8pt]
  |0; (\rep{1}, \rep{1})_{\bar{2}} \rangle  &&  |1; (\rep{1},  \rep{1},\rep{1})_{\bar{2}\bar{1}} \rangle &&  && |0; (\rep{1}, \rep{1}, \rep{6})_{\bar{2}0} \rangle &&  && |\bar{1}; (\rep{1}, \rep{1}, \rep{1})_{\bar{2}1} \rangle \\[8pt]
  |\bar{\frac{1}{2}}; (\rep{2}, \rep{1})_{\bar{1}} \rangle  &&  |\frac{1}{2}; (\rep{2}, \rep{1}, \rep{1})_{\bar{1}\bar{\frac{3}{2}}} \rangle && && |\bar{\frac{1}{2}}; (\rep{2}, \rep{1}, \rep{6})_{\bar{1}\bar{\frac{1}{2}}} \rangle && && |\bar{\frac{3}{2}}; (\rep{2}, \rep{1}, \rep{1})_{\bar{1}{\frac{1}{2}}} \rangle\\[8pt]
  |\bar{1}; (\rep{1}, \rep{1})_0 \rangle  &&  |0; (\rep{1}, \rep{1}, \rep{1})_{0\bar{2}} \rangle &&  && |\bar{1}; (\rep{1}, \rep{1}, \rep{6})_{0\bar{1}} \rangle && && |\bar{2}; (\rep{1}, \rep{1}, \rep{1})_{00} \rangle \\
  &&&&&&&&&\\
    &&&&&&&&&\\
 |\frac{1}{2}; (\rep{1}, \rep{2})_{\bar{1}} \rangle_\rho &&  && |1; (\rep{1}, \rep{2}, \rep{4})_{10} \rangle &&&& |0; (\rep{1}, \rep{2}, \overline{\rep{4}})_{11} \rangle &&  \\[8pt]
  |0; (\rep{2}, \rep{2})_{{0}} \rangle_\rho  &&   && |\frac{1}{2}; (\rep{2}, \rep{2}, \rep{4})_{\bar{2}\bar{\frac{1}{2}}} \rangle && && |\bar{\frac{1}{2}}; (\rep{2}, \rep{2}, \overline{\rep{4}})_{\bar{2}\frac{1}{2}} \rangle &&  \\[8pt]
  |\bar{\frac{1}{2}}; (\rep{1}, \rep{2})_{{1}} \rangle_\rho  &&  && |0; (\rep{1}, \rep{2}, \rep{4})_{\bar{1}\bar{1}} \rangle && && |\bar{1}; (\rep{1}, \rep{2}, \overline{\rep{4}})_{\bar{1}0} \rangle &&
\end{array}
\ee
Here the left $\nl'=2$ multiplet states carry $\mathfrak{so}(2)_l$ spacetime little group  and $\mathfrak{u}(2)_L\oplus\mathfrak{su}(2)_L$ R-symmetry plus enhanced flavour representations. The right $\nr'=0$ multiplet states carry $\mathfrak{so}(2)_r$ spacetime little group and $\mathfrak{su}(4)_R$ representations, where the $\mathfrak{su}(4)_R$ can be regarded as the remnant of the $\nr=4$ R-symmetry. The $\N_-=2$ supergravity and vector multiplet states carry $\mathfrak{so}(2)_{st}\oplus \mathfrak{u}(2)_L \oplus [\mathfrak{su}(2)_L\oplus\mathfrak{su}(4)_R]\oplus \mathfrak{u}(1)$ representations, where the spacetime helicity group $\mathfrak{so}(2)_{st}$ and the additional $\mathfrak{u}(1)$  are given by the sum and difference of the $\mathfrak{so}(2)_l$ and $\mathfrak{so}(2)_r$ generators, respectively. The charges carried by the extra $\mathfrak{u}(1)$ are given by the second subscript.

The $\mathfrak{u}(2)_L$ R-symmetry of the of the left multiplet carries over as the R-symmetry of the gravity plus vector multiplets. The additional $\mathfrak{u}(1)$ and   enhanced flavour $\mathfrak{su}(2)_L$ together with the  $\mathfrak{su}(4)_R$ R-symmetry remnant of the right multiplet are enhanced to provide the $\mathfrak{su}(6)$ isotropy group. As for the $\N_+=6$ twin we must take the same  linear combination of $\mathfrak{u}(1)$ generators to organise the states into $\mathfrak{su}(6)$ representations. Note, the R-symmetry representations simply go along for the ride. Using 
\be
\begin{array}{llllll}
\mathfrak{su}(6)&\supset& \mathfrak{su}(2)\oplus\mathfrak{su}(4)\oplus\mathfrak{u}(1) \\
&&\\
\rep{15}&\rightarrow&\rep{(1,1)}_{-4}+\rep{(2, 4)}_{-1}+\rep{(1, 6)}_{2}\\[5pt]
\rep{\overline{15}}&\rightarrow& \rep{(1,1)}_{4}+\rep{\left(2, \bar{4}\right)}_{1}+\rep{(1, 6)}_{-2}\\[5pt]
\end{array}
\ee
we find that the spectrum of $\eqref{n2}$ is reproduced. 

 This summarises the origin of the $D=4, (6,2)$ twins from  the perspective of Yang-Mills squared. It should be noted that the magic supergravity described here was previously double-copy constructed in \cite{Chiodaroli:2015wal} as the  the product of an $\N=2$ vector multiplet and $\N=0$ vector potential coupled to six adjoint scalars and 8 pseudo-real fermions in the $(\rep{2,8})$ of $\mathfrak{su}(2)\oplus\mathfrak{su}(4)$.

\subsection{Summary: the pyramid twins in $D=3, 4, 5, 6$}\label{summary}

The remaining examples in $D=3,4,5$ follow precisely the same pattern and we accordingly omit the details. The results are summarised in \autoref{PYR}, \autoref{D3},  \autoref{D4} and  \autoref{D5}. At this stage some comments are in order.  First, the twin relation generates new double-copy constructions from old. For example, as far as we are aware, the $D=4, \N_-=1$ twins  have not appeared as double-copies previously. In particular, for $\nl=1$ and $\nr=1$ we obtain  $\N_+=2$  supergravity minimally coupled to a single hypermultiplet with scalar coset $\Un(1,2)/\Un(2)$, which was double-copy constructed in \cite{Chiodaroli:2015wal}, but its twin, $\N_-=1$  supergravity minimally coupled to a single vector multiplet and two chiral multiplets, has not yet  appeared and remains to be tested at loop level.  There is in fact a two parameter family of $(2,1)$ twins coupled to vector and hyper multiplets \cite{Duff:2010ss}, which do not belong to the pyramid, but can be double-copy

\begin{table}
  \[
 \hspace{-0.25in}\xymatrix{
  D=6 &&&{\begin{array}{c}\frac{\SO(5, 5)}{\SO(5)\times\SO(5)}\\[5pt](2,2)\end{array}}\ar[r] \ar[dl]\ar@{--}@[red][d]&{\begin{array}{c}\frac{\SU^\star(4)}{\USp(2)}\\[5pt]((2,1), (0,1))\end{array}}\ar[dl]\ar@{-->}[dd]\\
    &&{\begin{array}{c}\frac{\SU^\star(4)}{\USp(2)}\\[5pt]((2,1), (0,1))\end{array}}\ar[r] \ar@{-->}[dd] &{\begin{array}{c}\frac{\Orth(1,1)\times \Sp(1)^2}{\Un(1)^2}\\[5pt]((1,1), (0, 1))\ar@{-->}@[red][d]\end{array}}\\
  D=5 &&&{\begin{array}{c}\frac{E_{6(6)}}{\USp(4)}\\[5pt](8)\end{array}}\ar[r]\ar[dl] \ar@{--}@[red][d]&{\begin{array}{c}\frac{\SU^\star(6)}{\USp(3)}\\[5pt](6, 2)\end{array}}\ar[dl]\ar@{-->}[dd]\\
    &&{\begin{array}{c}\frac{\SU^\star(6)}{\USp(3)}\\[5pt](6, 2)\end{array}}\ar[r] \ar@{-->}[dd]&{\begin{array}{c}\frac{\SO(1,1)\times \SO(5,1)}{\USp(2)}\\[5pt](4, 2)\end{array}}\ar@{-->}@[red][d]\\
  D=4 &&&{\begin{array}{c}\frac{E_{7(7)}}{\SU(8)}\\[5pt](8)\end{array}}\ar[r] \ar[dl]\ar@{--}@[red][d]&{\begin{array}{c}\frac{\SO^\star(12)}{\Un(6)}\\[5pt](6, 2)\end{array}}\ar[r] \ar[dl]&{\begin{array}{c}\frac{\SU(5,1)}{\Un(5)}\\[5pt](5, 1)\end{array}}\ar[dl] \ar@{-->}[ddd]\\
    &&{\begin{array}{c}\frac{\SO^\star(12)}{\Un(6)}\\[5pt](6, 2)\end{array}}\ar[r] \ar[dl]&{\begin{array}{c}\frac{\SU(1,1)\times\SO(6, 2)}{\Un(1)\times\Un(4)}\\[5pt](4, 2)\end{array}}\ar[r] \ar[dl]\ar@{--}@[red][d]& {\begin{array}{c}\frac{\SU(3,1)}{\Un(3)}\\[5pt](3, 2, 1)\end{array}}\ar[dl]\\
    &{\begin{array}{c}\frac{\SU(5,1)}{\Un(5)}\\[5pt](5, 1)\end{array}}\ar[r]  \ar@{-->}[ddd]& {\begin{array}{c}\frac{\SU(3,1)}{\Un(3)}\\[5pt](3, 2, 1)\end{array}} \ar[r] &{\begin{array}{c}\frac{\SU(2,1)}{\Un(2)}\\[5pt](2, 1)\end{array}}\ar@{-->}@[red][d] \\
  D=3  &&&{\begin{array}{c}\frac{E_{8(8)}}{\SO(16)}\\[5pt](16)\end{array}}\ar[r] \ar[dl]&{\begin{array}{c}\frac{E_{7(-5)}}{\SO(3)\times\SO(12)}\\[5pt](12, 4)\end{array}}\ar[r] \ar[dl]&{\begin{array}{c}\frac{E_{6(-14)}}{\Un(1)\times\SO(10)}\\[5pt](10, 2)\end{array}}\ar[r] \ar[dl]&{\begin{array}{c}\frac{F_{4(-20)}}{\SO(9)}\\[5pt](9, 1)\end{array}}\ar[dl] \\
    &&{\begin{array}{c}\frac{E_{7(-5)}}{\SO(3)\times\SO(12)}\\[5pt](12, 4)\end{array}}\ar[r] \ar[dl]&{\begin{array}{c}\frac{\SO(8,4)}{\SO(8)\times\SO(4)}\\[5pt](8, 4)\end{array}}\ar[r] \ar[dl]&{\begin{array}{c}\frac{\SU(4,2)}{\Un(4)\times\SU(2)}\\[5pt](6, 4,  2)\end{array}}\ar[r] \ar[dl]&{\begin{array}{c}\frac{\USp(2,1)}{\USp(2)\times\SU(2)}\\[5pt](5, 1)\end{array}}\ar[dl]\\
    &{\begin{array}{c}\frac{E_{6(-14)}}{\Un(1)\times\SO(10)}\\[5pt](10, 2)\end{array}}\ar[r] \ar[dl]&{\begin{array}{c}\frac{\SU(4,2)}{\Un(4)\times\SU(2)}\\[5pt](6, 4, 2)\end{array}}\ar[r] \ar[dl]&{\begin{array}{c}\frac{\SU(2,1)^2}{\Un(2)^2}\\[5pt](4, 2)\end{array}}\ar[r] \ar[dl]&{\begin{array}{c}\frac{\SU(2,1)}{\Un(2)}\\[5pt](3, 1)\end{array}}\ar[dl] \\
    {\begin{array}{c}\frac{F_{4(-20)}}{\SO(9)}\\[5pt](9, 1)\end{array}} \ar[r] &{\begin{array}{c}\frac{\USp(2,1)}{\USp(2)\times\SU(2)}\\[5pt](5, 1)\end{array}}\ar[r] &{\begin{array}{c}\frac{\SU(2,1)}{\Un(2)}\\[5pt](3, 1)\end{array}}\ar[r] &{\begin{array}{c}\frac{\SL(2, \R)}{\SO(2)}\\[5pt](2, 1)\end{array}} }
\]
\caption{ Pyramid of twin supergravities generated by the product of left and right super Yang-Mills theories in $D=3,4,5,6$. Each level is related by dimensional reduction as indicated by the vertical arrows. The horizontal arrows indicate consistent truncations effected by truncating the left or right Yang-Mills multiplets. The twins and triplets are indicated by $(\N_+, \N_-)$ and $(\N_+, \N^{+}_{-}, \N^{-}_{-})$, respectively, together with their common scalar manifolds. All such supergravity theories have a twin related by their left/right factors except for the  maximal cases along the ``exceptional spine'' highlighted in red. Consequently, for $D>6$ there  are no twin theories and this portion of the pyramid is omitted. Note,  $D=3$ is the exception to the exceptions in that maximal $\N=16$ supergravity does have a `trivial' $\N=1$ twin, but it is not obtained from our double-copy procedure and so is excluded.}\label{PYR}
\end{table}

\clearpage
\begin{turnpage}
\begin{table}
\caption{\label{D3} The twin supergravities in $D=3$. Here we give the left and right (super) Yang-Mills products yielding the twin $(\N_+, \N_-)$ supergravities. }
\begin{ruledtabular}
\begin{tabular}{ccc|ccc|cccccccccccccccccccc}
&&&&&&&&&&\\
 \multicolumn{3}{c|}{Left Yang-Mills-matter}   &   \multicolumn{3}{c|}{Right  Yang-Mills-matter} &        \multicolumn{4}{c}{Twin supergravities} \\
&&&&&&&&\\
  $\nl' $& Content & Symmetry & $\nr{}^{(\prime)} $& Content & Symmetry &  $\N_\pm $& Content & Symmetry &Coset\\
&&&&&&&&\\

\hline
&&&&&&&&\\
\multirow{2}{*}{ $4$}& \multirow{2}{*}{ $\bV_4\oplus\bC^{\rho}_{4}$} & \multirow{2}{*}{$\mathfrak{so}(4)_R\oplus\mathfrak{so}(3)_f$} & $8$&  $\bV_8$ & \multirow{2}{*}{ $\mathfrak{so}(7)$}   & $12$&  $\bG_{12}$&$\mathfrak{so}(12)_R\oplus\mathfrak{so}(3)$& \multirow{2}{*}{ $\frac{E_{7(-5)}}{\SO(12)\times\SO(3)}$}\\
  &  &  & 0& $\tA \rep{(1)} \oplus\tX^{\tilde{\rho}} \rep{(8)}\oplus \tphi \rep{(7)}  $  & & 4&    $\bG_4\oplus 16\bV_4$&$\mathfrak{so}(4)_R\oplus\mathfrak{so}(12)_\text{Isotropy}\oplus\mathfrak{so}(3)$&\\
&&&&&&&&\\
\hline

&&&&&&&&\\
\multirow{2}{*}{ $4$}& \multirow{2}{*}{ $\bV_4\oplus\bC^{\rho}_{4}$} & \multirow{2}{*}{$\mathfrak{so}(4)_R\oplus\mathfrak{so}(3)_f$} & $4$&  $\bV_4$ &\multirow{2}{*}{  $\mathfrak{so}(4)$  } & $8$&  $\bG_8\oplus4\bV_8$&$\mathfrak{so}(8)_R\oplus\mathfrak{so}(4)_\text{Isotropy}$& \multirow{2}{*}{ $\frac{\SO(8, 4)}{\SO(8)\times\SO(4)}$}\\
  &  &  &0&  $\tA \rep{(1,1)} \oplus\tX^{\tilde{\rho}} \rep{(2,2)}\oplus \tphi \rep{(3,1)}  $  &  & 4&    $\bG_4\oplus 8\bV_4 $&$\mathfrak{so}(4)_R\oplus\mathfrak{so}(8)_\text{Isotropy}\oplus\mathfrak{so}(4)$&\\
&&&&&&&&\\
\hline

&&&&&&&&\\
\multirow{2}{*}{ $2$}& \multirow{2}{*}{ $\bV_2\oplus\bC^{\rho}_{2}$} & \multirow{2}{*}{$\mathfrak{so}(2)_r\oplus\mathfrak{so}(2)_f$} & $8$&  $\bV_8$ & \multirow{2}{*}{ $\mathfrak{so}(7)$}   & $10$&  $\bG_{10}$&$\mathfrak{so}(10)_R\oplus\mathfrak{so}(2)$& \multirow{2}{*}{ $\frac{E_{6(-14)}}{\SO(10)\times\Un(1)}$}\\
  &  &  &0&  $\tA \rep{(1)} \oplus\tX^{\tilde{\rho}} \rep{(8)}\oplus \tphi \rep{(7)}  $   &  & 2&    $\bG_2\oplus \bV_2\oplus10\bV_2\oplus 5\bC_{2}$&$\mathfrak{so}(2)_R\oplus \mathfrak{so}(10)_\text{Isotropy}\oplus\mathfrak{so}(2)$&\\
&&&&&&&&\\
\hline

&&&&&&&&\\
\multirow{2}{*}{ $2$}& \multirow{2}{*}{ $\bV_2\oplus\bC^{\rho}_{2}$} & \multirow{2}{*}{$\mathfrak{so}(2)_r\oplus\mathfrak{so}(2)_f$}& $4$&  $\bV_4$ & \multirow{2}{*}{  $\mathfrak{so}(4)$  }  & $6$&  $\bG_6\oplus 2 \bV_6$&$\mathfrak{so}(6)_R\oplus\mathfrak{so}(3)\oplus\mathfrak{so}(2)_{\text{Isotropy}}$& \multirow{2}{*}{ $\frac{\SU(4,2)}{\SU(4)\times\Un(2)}$}\\
  &  &  &0&  $\tA \rep{(1,1)} \oplus\tX^{\tilde{\rho}} \rep{(2,2)}\oplus \tphi \rep{(3,1)}  $   &  & 2&    $\bG_2\oplus\bV_2\oplus 4 \bV_2\oplus 3\bC_{2}$&$\mathfrak{so}(2)_R\oplus[\mathfrak{su}(4)\oplus\mathfrak{u}(2)]_\text{Isotropy}$&\\
&&&&&&&&\\

\hline

&&&&&&&&\\
\multirow{2}{*}{ $2$}& \multirow{2}{*}{ $\bV_2\oplus\bC^{\rho}_{2}$} & \multirow{2}{*}{$\mathfrak{so}(2)_r\oplus\mathfrak{so}(2)_f$}& $2$&  $\bV_2$ & \multirow{2}{*}{  $\mathfrak{so}(2)$  }   & $4$&  $\bG_4\oplus\bV_4\oplus\bC_{4}$&$\mathfrak{so}(4)_R\oplus\mathfrak{so}(2)\oplus\mathfrak{so}(2)_{\text{Isotropy}}$& \multirow{2}{*}{ $\frac{\SU(2,1)}{\Un(2)}\times\frac{\SU(2,1)}{\Un(2)}$}\\
  &  &  &0&  $\tA \oplus2\tX^{\tilde{\rho}}\oplus\phi   $  &  & 2&    $\bG_2\oplus \bV_2\oplus\bV_2\oplus 2\bC_{2}$&$\mathfrak{so}(2)_R\oplus[\mathfrak{u}(2)\oplus\mathfrak{u}(2)]_\text{Isotropy}$&\\
&&&&&&&&\\

\hline
&&&&&&&&\\
\multirow{2}{*}{ $1$}& \multirow{2}{*}{ $\bV_1\oplus\bC^{\rho}_{1}$} & \multirow{2}{*}{$\varnothing$} & $8$&  $\bV_8$ & \multirow{2}{*}{  $\mathfrak{so}(7)$  }   & $9$&  $\bG_9$&$\mathfrak{so}(9)_R$& \multirow{2}{*}{ $\frac{F_{4(-20)}}{\SO(9)}$}\\
  &  &  &0&  $\tA \rep{(1)} \oplus\tX^{\tilde{\rho}} \rep{(8)}\oplus \tphi \rep{(7)}  $  & & 1&    $\bG_1\oplus 16\bV_1$&$\mathfrak{so}(9)_\text{Isotropy}$&\\
&&&&&&&&\\
\hline

&&&&&&&&\\
\multirow{2}{*}{ $1$}& \multirow{2}{*}{ $\bV_1\oplus\bC^{\rho}_{1}$} & \multirow{2}{*}{$\varnothing$} & $4$&  $\bV_4$ &\multirow{2}{*}{  $\mathfrak{so}(4)$  }   & $5$&  $\bG_5\oplus\bV_5$&$\mathfrak{so}(5)_R\oplus\mathfrak{so}(3)$& \multirow{2}{*}{ $\frac{\USp(2, 1)}{\USp(2)\times\SU(2)}$}\\
  &  &  &0&  $\tA \rep{(1,1)} \oplus\tX^{\tilde{\rho}} \rep{(2,2)}\oplus \tphi \rep{(3,1)}  $  &  & 1&    $\bG_1\oplus 8\bV_1$&$\mathfrak{so}(5)_\text{Isotropy}\oplus\mathfrak{so}(3)$&\\
&&&&&&&&\\
\hline

&&&&&&&&\\
\multirow{2}{*}{ $1$}& \multirow{2}{*}{ $\bV_1\oplus\bC^{\rho}_{1}$} & \multirow{2}{*}{$\varnothing$} & $2$&  $\bV_2$ & \multirow{2}{*}{  $\mathfrak{so}(2)$  }  & $3$&  $\bG_3\oplus\bV_3$&$\mathfrak{so}(3)_R\oplus\mathfrak{so}(2)$& \multirow{2}{*}{ $\frac{\SU(2,1)}{\Un(2)}$}\\
  &  &  &0&  $\tA \oplus2\tX^{\tilde{\rho}}\oplus\phi   $   &  & 1&    $\bG_1\oplus 4\bV_1$&$\mathfrak{so}(2)\oplus\mathfrak{so}(3)_\text{Isotropy}$&\\
&&&&&&&&\\
\hline

&&&&&&&&\\
\multirow{2}{*}{ $1$}& \multirow{2}{*}{ $\bV_1\oplus\bC^{\rho}_{1}$} & \multirow{2}{*}{$\varnothing$}& $1$&  $\bV_1$ & \multirow{2}{*}{  $\varnothing$}  & $2$&  $\bG_2\oplus\bV_2$&$\mathfrak{so}(2)_R$& \multirow{2}{*}{ $\frac{\SL(2, \R)}{\SO(2)}$}\\
  &  &  &0&  $\tA \oplus\tX^{\tilde{\rho}}  $  &  & 1&    $\bG_1\oplus 2\bV_1$&$\mathfrak{so}(2)_\text{Isotropy}$&\\
&&&&&&&&\\

\end{tabular}
\end{ruledtabular}
\end{table}
\end{turnpage}

\begin{turnpage}
\begin{table}
\caption{\label{D4} The twin supergravities in $D=4$. Here we give the left and right (super) Yang-Mills products yielding the twin $(\N_+, \N_-)$ supergravities. In the first column we summarise the left Yang-Mills-matter theories and their global symmetries, where $\bV_\nl$ and $\bC_\nl$ are in the adjoint and $\rho$ representations of the left gauge group $G$ respectively. In the second column we summarise the right Yang-Mills-matter theories before and after  \eqref{rdecomp} has been applied. Again, for both cases  their global symmetries are given and for the $\nr=0$ theories we have indicated the representation carried by each field (omitting all $\mathfrak{u}(1)$ charges). Note,  the  fermions of  the $\nr=0$ theories are always in the   $\tilde{\rho}$ representation of the right gauge group $\tG$, while the vectors and scalars remain in the adjoint. In the final column we have tabulated the resulting pairs of twin supergravity theories and their common scalar coset manifolds. Note, the final row can be generalised to an infinite sequence of $\N_+=2$ self-mirror minimally coupled supergravity theories and their $\N_-=1$ twins, as discussed in \autoref{othertwins}.}
\begin{ruledtabular}
\begin{tabular}{ccc|ccc|ccccccccccccccccccccc}
&&&&&&&&&&\\
 \multicolumn{3}{c|}{Left Yang-Mills-matter}   &   \multicolumn{3}{c|}{Right  Yang-Mills-matter} &        \multicolumn{4}{c}{Twin supergravities} \\
&&&&&&&&\\
  $\nl' $& Content & Symmetry  & $\nr{}^{(\prime)} $& Content & Symmetry   &  $\N_\pm $& Content & Symmetry &Coset\\
&&&&&&&&\\

\hline
&&&&&&&&\\
\multirow{2}{*}{ $2$}& \multirow{2}{*}{ $\bV_2\oplus\bH^{\rho}_{2}$} & \multirow{2}{*}{$\mathfrak{u}(2)_R\oplus\mathfrak{su}(2)_f$} & $4$&  $\bV_4$ & $\mathfrak{su}(4)_R$   & $6$&  $\bG_6$&$\mathfrak{u}(6)_R$& \multirow{2}{*}{ $\frac{\SO^\star(12)}{\Un(6)}$}\\
  &    &&$0$&  $\tA \rep{(1)} \oplus\tX^{\tilde{\rho}} \rep{(4)}\oplus \tphi \rep{(6)}  $  & $\mathfrak{su}(4)$ & 2&    $\bG_2\oplus \bV_2(\rep{15})$&$\mathfrak{u}(2)_R\oplus\mathfrak{u}(6)_\text{Isotropy}$&\\
&&&&&&&&\\
\hline

&&&&&&&&\\
\multirow{2}{*}{ $2$}& \multirow{2}{*}{ $\bV_2\oplus\bH^{\rho}_{2}$} & \multirow{2}{*}{$\mathfrak{u}(2)_R\oplus\mathfrak{su}(2)_f$} & $2$&  $\bV_2$ & $\mathfrak{u}(2)_R$  & $4$&  $\bG_4\oplus2\bV_4$&$\mathfrak{u}(4)_R\oplus\mathfrak{so}(2)_\text{Isotropy}$& \multirow{2}{*}{ $\frac{\SL(2, \R)\times\SO(6, 2)}{\SO(2)\times\Un(4)}$}\\
  &    &&$0$&  $\tA \rep{(1)} \oplus\tX^{\tilde{\rho}} \rep{(2)}\oplus 2\tphi \rep{(1)}  $  & $\mathfrak{u}(2)$ & 2&    $\bG_2\oplus \bV_2(\rep{1})\oplus \bV_2(\rep{6})$&$\mathfrak{u}(2)_R\oplus\mathfrak{u}(4)_\text{Isotropy}$&\\
&&&&&&&&\\
\hline

&&&&&&&&\\
\multirow{2}{*}{ $1$}& \multirow{2}{*}{ $\bV_1\oplus\bC^{\rho}_{1}$} & \multirow{2}{*}{$\mathfrak{u}(1)_R\oplus\mathfrak{u}(1)_f$}& $4$&  $\bV_4$ & $\mathfrak{su}(4)_R$ & $5$&  $\bG_5$&$\mathfrak{u}(5)_R$& \multirow{2}{*}{ $\frac{\SU(5,1)}{\Un(5)}$}\\
  &    &&$0$&  $\tA \rep{(1)} \oplus\tX^{\tilde{\rho}} \rep{(4)}\oplus \tphi \rep{(6)}  $   & $\mathfrak{su}(4)$ & 1&    $\bG_1\oplus \bV_1(\rep{10})\oplus \bC_1(\rep{5})$&$\mathfrak{u}(1)_R\oplus\mathfrak{u}(5)_\text{Isotropy}$&\\
&&&&&&&&\\
\hline

&&&&&&&&\\
\multirow{2}{*}{ $1$}& \multirow{2}{*}{ $\bV_1\oplus\bC^{\rho}_{1}$} & \multirow{2}{*}{$\mathfrak{u}(1)_R\oplus\mathfrak{u}(1)_f$}& $2$&  $\bV_2$ & $\mathfrak{u}(2)_R$ & $3$&  $\bG_3\oplus\bV_3$&$\mathfrak{u}(3)_R\oplus\mathfrak{u}(1)$& \multirow{2}{*}{ $\frac{\Un(3,1)}{\Un(3)\times\Un(1)}$}\\
  &    &&$1$&  $\tA \rep{(1)} \oplus\tX^{\tilde{\rho}} \rep{(2)}\oplus 2 \tphi \rep{(1)}  $  & $\mathfrak{u}(2)$ & 1&    $\bG_1\oplus \bV_1(\rep{1})\oplus \bV_1(\rep{3})\oplus \bC_1(\rep{3})$&$\mathfrak{u}(1)_R\oplus\mathfrak{u}(3)_\text{Isotropy}$&\\

&&&&&&&&\\

\hline

&&&&&&&&\\
\multirow{2}{*}{ $1$}& \multirow{2}{*}{ $\bV_1\oplus\bC^{\rho}_{1}$} & \multirow{2}{*}{$\mathfrak{u}(1)_R\oplus\mathfrak{u}(1)_f$}& $1$&  $\bV_1$ & $\mathfrak{u}(1)_R$   & $2$&  $\bG_2\oplus\bH_2$&$\mathfrak{u}(2)_R\oplus\mathfrak{u}(1)$& \multirow{2}{*}{ $\frac{\Un(2,1)}{\Un(2)\times\Un(1)}$}\\
  &    &&$0$&  $\tA \oplus\tX^{\tilde{\rho}}   $  & $\mathfrak{u}(1)$ & 1&    $\bG_1\oplus \bV_1(\rep{1})\oplus \bC_1(\rep{2})$&$\mathfrak{u}(1)_R\oplus\mathfrak{u}(2)_\text{Isotropy}$&\\
&&&&&&&&\\

\end{tabular}
\end{ruledtabular}
\end{table}
\end{turnpage}

\begin{turnpage}
\begin{table}
\caption{\label{D5} The twin supergravities in $D=5$. Here we give the left and right (super) Yang-Mills products yielding the twin $(\N_+, \N_-)$ supergravities. }
\begin{ruledtabular}
\begin{tabular}{ccc|ccc|ccccccccccccccccccccc}
&&&&&&&&&&\\
 \multicolumn{3}{c|}{Left Yang-Mills-matter}   &   \multicolumn{3}{c|}{Right  Yang-Mills-matter} &        \multicolumn{4}{c}{Twin supergravities} \\
&&&&&&&&\\
  $\nl' $& Content & Symmetry  & $\nr{}^{(\prime)} $& Content & Symmetry   &  $\N_\pm $& Content & Symmetry &Coset\\
&&&&&&&&\\

\hline
&&&&&&&&\\
\multirow{2}{*}{ $2$}& \multirow{2}{*}{ $\bV_2\oplus\bH^{\rho}_{2}$} & \multirow{2}{*}{$\mathfrak{sp}(1)_R\oplus \mathfrak{sp}(1)_f$} & $4$&  $\tbV_4$ & $\mathfrak{sp}(2)_R$   & $6$&  $\bG_6$&$\mathfrak{sp}(3)_R$& \multirow{2}{*}{ $\frac{\SU^*(6)}{\USp(3)}$}\\
  &    &&$0$&  $\tA \rep{(1)} \oplus\tX^{\tilde{\rho}} \rep{(4)}\oplus \tphi \rep{(5)}  $  & $\mathfrak{sp}(2)$ & 2&    $\bG_2\oplus \bV_2(\rep{14})$&$ \mathfrak{sp}(1)_R\oplus \mathfrak{sp}(3)_\text{Isotropy}$&\\
&&&&&&&&\\
\hline

&&&&&&&&\\
\multirow{2}{*}{ $2$}& \multirow{2}{*}{ $\bV_2\oplus\bH^{\rho}_{2}$} & \multirow{2}{*}{$\mathfrak{sp}(1)_R\oplus \mathfrak{sp}(1)_f$} & $2$&  $\tbV_2$ & $ \mathfrak{sp}(1)_R$  & $4$&  $\bG_4\oplus \bV_4$&$ \mathfrak{sp}(2)_R$& \multirow{2}{*}{ $\frac{\Orth(5,1)\times \Orth(1,1)}{\USp(2)}$}\\
  &    &&$0$&  $\tA \rep{(1)} \oplus\tX^{\tilde{\rho}} \rep{(2)}\oplus \tphi \rep{(1)}  $  & $ \mathfrak{sp}(1)$ & 2&    $\bG_2\oplus \bV_2(\rep{1})\oplus \bV_2(\rep{5})$&$ \mathfrak{sp}(1)_R\oplus \mathfrak{sp}(2)_\text{Isotropy}$&\\
&&&&&&&&\\

\end{tabular}
\end{ruledtabular}

\caption{\label{D6} The twin supergravities in $D=6$. The $\N_+$ twin is generated as a truncation of the parent as for $D=3,4,5$. The $\N_- $ twin requires an additional chirality flip   of the left Yang-Mills-matter multiplet.}
\begin{ruledtabular}
\begin{tabular}{ccc|ccc|ccccccccccccccccccccc}
&&&&&&&&&&\\
 \multicolumn{3}{c|}{Left Yang-Mills-matter}   &   \multicolumn{3}{c|}{Right  Yang-Mills-matter} &        \multicolumn{4}{c}{Twin supergravities} \\
&&&&&&&&\\
  $\nl' $& Content & Symmetry  & $\nr{}^{(\prime)} $& Content & Symmetry   &  $\N_\pm $& Content & Symmetry &Coset\\
&&&&&&&&\\

\hline
&&&&&&&&\\
$(1,0)$& $\bV_{1,0}\oplus\bH^{\rho}_{1,0}$ & $\mathfrak{sp}(1)_R\oplus \mathfrak{sp}(1)_f$ & $(1,1)$&  $\tbV_{1,1}$ & $\mathfrak{sp}(1)_R\oplus\mathfrak{sp}(1)_R$   & $(2,1)$&  $\bG_{2,1}$&$\mathfrak{sp}(2)_R\oplus \mathfrak{sp}(1)_R$& \multirow{2}{*}{ $\frac{\SU^\star(4)\times \Sp(1)}{\USp(2)\times\Un(1)}$}\\
$(0,1)$  & $\bV_{0,1}\oplus\bH^\rho_{0,1}$  &$\mathfrak{sp}(1)_f\oplus\mathfrak{sp}(1)_R $& $(0,0)$&  $\tA(\rep{1})\oplus\tX_{-}^{\rho}(\rep{2})\oplus\tX_{+}^{\rho}(\rep{2})\oplus\tilde{\phi}(\rep{4}) $  & $\mathfrak{sp}(1)_R$ & $(0,1)$&    $\bG_{0,1}\oplus \bV_{0,1}(\rep{4}+\rep{4})\oplus\bT_{0,1}(\rep{5})$&$\mathfrak{sp}(1)_R\oplus\mathfrak{sp}(2)_\text{Isotropy}$&\\
&&&&&&&&\\
\hline

&&&&&&&&\\
$(1,0)$&  $\bV_{1,0}\oplus\bH^{\rho}_{1,0}$ & $\mathfrak{sp}(1)_R\oplus \mathfrak{sp}(1)_f$ & $(1,0)$&  $\tbV_{1,0}$ & $\mathfrak{sp}(1)_R$  & $(2,0)$&  $\bG_{2,0}\oplus\bT_{2,0}$&$\mathfrak{sp}(2)_R$& \multirow{2}{*}{ $\frac{\SU^*(4)}{\USp(2)}$}\\
 $(0,1)$ & $\bV_{0,1}\oplus\bH^{\rho}_{0,1}$ &$ \mathfrak{sp}(1)_f\oplus\mathfrak{sp}(1)_R$& $(0,0)$&  $\tA(\rep{1})\oplus\tX_{-}^{\rho}  (\rep{2})$  & $\mathfrak{sp}(1)$ & $(0,1)$&    $\bG_{0,1}\oplus \bT_{0,1}(\rep{5})$&$\mathfrak{sp}(1)_R\oplus\mathfrak{sp}(2)_\text{Isotropy}$&\\
&&&&&&&&\\
\hline

&&&&&&&&\\
$(1,0)$& $\bV_{1,0}\oplus\bH^{\rho}_{1,0}$ & $\mathfrak{sp}(1)_R\oplus \mathfrak{sp}(1)_f$ & $(0,1)$&  $\tbV_{0,1}$ & $\mathfrak{sp}(1)_R$ & $(1,1)$&  $\bG_{1,1}$&$\mathfrak{sp}(1)_R\oplus \mathfrak{sp}(1)_R$& \multirow{2}{*}{ $\frac{\Orth(1,1)\times \Sp(1)^2}{\Un(1)^2}$}\\
 $(0,1)$ & $\bV_{0,1}\oplus\bH^\rho_{0,1}$ &$\mathfrak{sp}(1)_f\oplus\mathfrak{sp}(1)_R$& $(0,0)$&  $\tA \rep{(1)} \oplus\tX_{+}^{\tilde{\rho}} \rep{(2)}  $   & $\mathfrak{sp}(1)$ & $(0,1)$&    $\bG_{0,1}\oplus \bV_{0,1}(\rep{4})\oplus \bT_{0,1}(\rep{1})$&$\mathfrak{sp}(1)_R\oplus\mathfrak{sp}(1)_\text{Isotropy}$&\\
&&&&&&&&\\

\end{tabular}
\end{ruledtabular}

\end{table}

\end{turnpage}

\clearpage
\global\pdfpageattr\expandafter{\the\pdfpageattr/Rotate 90}
\clearpage
\global\pdfpageattr\expandafter{\the\pdfpageattr/Rotate 0}
\noindent constructed as discussed in \autoref{othertwins}. Note, the associated sequence of special K\"ahler symmetric scalar manifolds appearing in the $\N_+=2$ theories can also be coupled to $\N_-=1, D=4$ supergravity because their kinetic vector matrices are holomorphic \cite{Andrianopoli:2007rm}.   %Their double-copy construction will be treated in a follow-up work \cite{Anastasiou:2016alternatives}.

Second, our approach applied to the prototypical $D=4, (6,2)$ twin pair gives an alternative double-copy construction  of the quaternionic magic $D=4, \N=2$ supergravity, which was previously obtained in \cite{Chiodaroli:2015wal} using a different pair of Yang-Mills-matter factors. This serves to highlight a general  feature of the double-copy construction for matter-coupled supergravities: the factorisation into left and right (super) Yang-Mills multiplets is not necessarily unique. The $D=4, (4, 2)$ twin pair is a clear example. The $\N_+=4$ supergravity comes coupled to two vector multiplets and 
 follows from the product of two, $\nl=\nr=2$, super Yang-Mills multiplets. As a truncation of the parent $\N=6$ theory it is schematically given by, 
\be\label{con}
[\bV_4]\otimes[\tbV_2]=\bG_6\longrightarrow [\bV_2\oplus\bC^{\rho}_2]\otimes[\tbV_2]= \bG_4 \oplus 2\bV_4.
\ee
Its twin $\N_-=2$ supergravity is coupled to seven vector multiplets and follows  from  the same principle applied to $\tbV_2$,
\be\begin{split}
[\bV_2\oplus\bC^{\rho}_2]\otimes[\tbV_2]\longrightarrow [\bV_2\oplus\bC^{\rho}_2]\otimes[\tA\oplus \tX_a^{\tilde{\rho}}\oplus\tphi]& = [\bV_2\otimes \tA]\oplus[\bC^{\rho}_2\otimes\tX_a^{\tilde{\rho}} ]\oplus[\bV_2 \oplus  \tphi] \\&=[\bG_2 \oplus \bV_2]  \oplus [2\bV_{2a}]\oplus[2\bV_{2}],
\end{split}
\ee
where $a=1,2$ is an $\mathfrak{su}(2)$  index, the remnant $\nr=2$ R-symmetry, and $\tphi$ is a complex scalar. The common scalar coset is given by $\SL(2, \R)\times \SO(6, 2)/\Un(1)\times\Un(4)$. The $\Un(4)$ is the R-symmetry of the $\N_+=4$  theory and the matter isotropy group of the $\N_-=2$ theory, rotating the six vector multiplets with a matter$\otimes$matter origin. Both theories, however, admit an alternative construction \cite{Anastasiou:2013hba},
\be\label{alt}
\bG_4 \oplus 2\bV_4 = \bV_4\otimes[\tA \oplus  2 \tphi]
\quad
\text{and} 
\quad
\bG_2 \oplus  7\bV_2 = \bV_2\otimes[\tA \oplus  6 \tphi],
\ee
where the $n$ scalar fields of the right multiplets are required to transform in the vector representation of $\SO(n)$. It turns out there is a plethora of non-unique decompositions of this type. The full classification  of all supergravities admitting more than one  Yang-Mills factorisations will be  given in \cite{Anastasiou:2016alternatives}.

The $D=6$ case, given in \autoref{D6},  is a little more subtle. In particular, we must take to account the possible chiralities, $\N=(n, m)$.  The big twin is obtained as a truncation of its parent following the prescription laid out above. To obtain the little twin, however,  the decomposition of the right Yang-Mills multiplet must be accompanied by a flip of the chirality of the left Yang-Mills-matter multiplet:
\be
\begin{array}{lllllll}
\text{Parent} &\bV_{(n, m)}&\otimes& \tbV_{(\tilde{n}, \tilde{m})}&=&\bG_{(n+\tilde{n}, m+\tilde{m})}\oplus\bM_{(n+\tilde{n}, m+\tilde{m})} \\[8pt]
\text{Big twin}&\bV_{(n', m')}\oplus \bH^{\rho}_{(n', m')}&\otimes& \tbV_{(\tilde{n}, \tilde{m})}&=&\bG_{(n_+, m_+)}\oplus\bM_{(n_+, m_+)} \\[8pt]
\text{Little twin} &\underbrace{\bV_{(m', n')}\oplus  \bH^{\rho}_{(m', n')}}_{\text{Chirality flipped}}&\otimes& \tbV_{(\tilde{n}', \tilde{m}')}\oplus \tbC^{\rho}_{(\tilde{n}', \tilde{m}')}\oplus \cdots&=&\bG_{(n_-, m_-)}\oplus\bM_{(n_-, m_-)} \\[8pt]
\end{array}
\ee
For example, the $(\N_+, \N_-)=((2,1), (0,1))$ is given by
\be
\begin{array}{lllllll}
\text{Parent} &\bV_{(1, 1)}&\otimes& \tbV_{(1, 1)}&=&\bG_{(2, 2)} \\[8pt]
\text{Big twin}&\bV_{(1, 0)}\oplus\bH^{\rho}_{(1, 0)}&\otimes& \tbV_{(1,1)}&=&\bG_{(2, 1)} \\[8pt]
\text{Little twin} &{\bV_{(0, 1)}\oplus\bH^{\rho}_{(0, 1)}}&\otimes& \tA\oplus2(\tX_{-}^{\rho}, \tX_{+}^{\rho})\oplus4\tilde{\phi} &=&\bG_{(0, 1)}\oplus8\bV_{(0, 1)} \oplus5\bV_{(0, 1)} \\[8pt]
\end{array}
\ee
All  three cases are presented in \autoref{D6}. 

Note, the  $(\N_+, \N_-)=((2,1), (0,1))$ example can also be generated by using tensor multiplets $\bT_{(n,m)}$, at least at the level of free on-shell states:
\be
\begin{array}{lllllll}
\text{Parent} &\bT_{(0, 2)}&\otimes& \tbT_{(2, 0)}&=&\bG_{(2, 2)} \\[8pt]
\text{Big twin}&\bT_{(0, 1)}\oplus\bH^{\rho}_{(0, 1)}&\otimes& \tbT_{(2,0)}&=&\bG_{(2, 1)} \\[8pt]
\text{Little twin} &\bT_{(0, 1)}\oplus\bH^{\rho}_{(0, 1)}&\otimes& \tilde{B}\oplus 4 \tX_{+}^{\rho}\oplus5\tilde{\phi} &=&\bG_{(0, 1)}\oplus8\bV_{(0, 1)} \oplus5\bT_{(0, 1)} \\[8pt]
\end{array}
\ee
The  $D=6,$ $\bG_{(2,2)}$ multiplet is the unique maximally supersymmetric gravity multiplet that admits two factorisations,  
\be\label{2,2}
\bG_{(2,2)}=\bT_{(2, 0)}\otimes \tbT_{(0, 2)}, \qquad \bG_{(2,2)}=\bV_{(1, 1)}\otimes \tbV_{(1, 1)}.
\ee 
Although the $D=6, (2,0)$ theories  are  intrinsically `strongly coupled' and do not admit any conventional Lagrangian description,   the existence of well-defined asymptotic states facilitates a direct analysis of the S-matrix  \cite{Huang:2010rn}. The tree-level amplitudes may be defined as the purely pole part of the S-matrix, although strong coupling implies that they cannot be interpreted as the leading term in any perturbative expansion. In the conformal phase there are no non-vanishing tree-level amplitudes for the self-dual tensor that respect the $(2,0)$  super-Poincar\'e invariance \cite{Huang:2010rn, Czech:2011dk}, leaving  the double-copy origin of the $\bG_{(2, 2)}$ amplitudes mysterious from this perspective.  One approach is to consider M5-branes on $\R^{1,4}\times S^1$ with self-dual strings (M2-branes ending on the M5-branes) wrapping the $S^1$ to give a tower of massive Kaluza-Klein modes in five dimensions. Then there is a non-trivial three-point amplitude for the self-dual tensor, which squares to give an  amplitude of the $D=6, (4,0)$ theory on $\R^{1,4}\times S^1$ \cite{Czech:2011dk}. In the massless limit the self-dual tensor amplitude reduces to that of  $D=5, \N=4$ super Yang-Mills  so that its square  correctly produces the corresponding   $D=5, \N=8$ supergravity amplitude \cite{Czech:2011dk}. This is consistent with the observation that in the linear approximation the $(4,0)$ theory dimensionally reduced on a circle is $D=5, \N=8$ supergravity  \cite{Hull:2000zn, Hull:2000rr}. Alternatively, the product of the $(2,0)$ and $(0, 2)$ amplitudes  gives that of $D=6, \N=8$ supergravity on $\R^{1,4}\times S^1$, as suggested by \eqref{2,2}, which  also gives the $D=5, \N=8$ supergravity amplitude in the massless limit.

Finally, a comment on $D=3, \N=1$ theories is needed. As  noted in \cite{Roest:2009sn} and \autoref{twins} all $\N_+>1$ theories (assuming all vectors have been dualised to scalars) in $D=3$  have an $\N_-=1$ twin. This follows from the fact that $D=3, \N=1$ supergravity can be coupled to scalars parametrising  any Riemannian    manifold and all admissible scalar cosets for $\N>1$ are Riemannian. For this reason such twins are typically regarded as trivial. However, for the pyramid of twins, \autoref{PYR}, they are natural in the sense that they follow from the same double-copy construction  described.  Note however, the $D=3, (16, 1)$ twin pair is \emph{not} obtained in this manner and, as such, it should be regarded as belonging to the excluded maximal spine.  It is also excluded on the basis that the $D=3, (16, 1)$ twins do  not have a  parent supergravity.  

\subsection{The triplets}\label{triplets}

In four dimensions there is a $(\N_+, \N^{+}_{-}, \N^{-}_{-})=(3, 2, 1)$ triplet of supergravity theories, which descends to a $(6, 4, 2)$ triplet in $D=3$. The notation $\N^{\pm}_{-}$ is used to indicate that $\N_+$ is the big sibling of both $\N^{\pm}_{-}$, while  $\N^{-}_{-}$ is the little sibling of $\N^{+}_{-}$. The common scalar manifold is 
\be
\frac{\SU(3,1)}{\SU(3)\times\Un(1)}.
\ee

We find that the two sub-twins $(\N_+, \N^{+}_{-})$ and $(\N_+, \N^{-}_{-})$ follow from the same considerations as above, as the $\N_+=3$ theory belongs to the pyramid. Specifically, for the $(3, 2)$ pair we have an $\N=5$ parent,
\be
\begin{array}{llllllllllllll}\label{32pair}
\text{Parent} &\bV_{4}&\otimes& \tbV_{1}&=&\bG_{5} \\[8pt]
\text{Big twin}&\bV_{2}\oplus\bH^{\rho}_{2}&\otimes& \tbV_{1}&=& \bG_{3}\oplus\bV_3 \\[8pt]
\text{Little twin} &\bV_{2}\oplus\bH^{\rho}_{2}&\otimes& \tA\oplus\tX^{\rho} &=& \bG_{2}\oplus3\bV_{2} 
\end{array}
\ee
The left and right symmetries of the big twin factors are
\be
\mathfrak{so}(2)_l\oplus \mathfrak{u}(2)_R\oplus\mathfrak{su}(2)_f\quad\text{and}\quad \mathfrak{so}(2)_{r}\oplus \mathfrak{u}(1)_R,
\ee
where $\mathfrak{u}(2)_R$ is the left $\N'=2$ R-symmetry and $\mathfrak{su}(2)_f$ is a remnant of the $\N=4$ R-symmetry feeding into the $\N=5$ parent. They sit inside  the $\N=3$ algebra as
\be
[\mathfrak{u}(2)_R\oplus\mathfrak{u}(1)_R]\oplus\mathfrak{u}(1)\subset \mathfrak{u}(3)_R\oplus\mathfrak{u}(1)_{\text{Isotropy}},
\ee
where the additional $\mathfrak{u}(1)$  is given by the difference of the $\mathfrak{so}(2)_l$ and $\mathfrak{so}(2)_r$ generators as usual. Note, the $\mathfrak{su}(2)_f$ acts trivially on all gravitational states as it only acts non-trivially on the $\bH^{\rho}_{2}$ multiplet, which plays no role here. 

Similarly, the left and right symmetries of the little twin factors are
\be
\mathfrak{so}(2)_l\oplus \mathfrak{u}(2)_R\oplus\mathfrak{su}(2)_f\quad\text{and}\quad \mathfrak{so}(2)_r\oplus \mathfrak{u}(1)_f,
\ee
where the right $\mathfrak{u}(1)_f$ is now  a remnant  of the right $\nr=1$ R-symmetry. They sit inside  the $\N=2$ algebra as
\be
\mathfrak{u}(2)_R\oplus[\mathfrak{su}(2)_f\oplus\mathfrak{u}(1)_f]\oplus\mathfrak{u}(1)\subset \mathfrak{u}(2)_R\oplus [\mathfrak{su}(3)\oplus\mathfrak{u}(1)]_{\text{Isotropy}}
\ee
where again the additional $\mathfrak{u}(1)$  is given by the difference of the $\mathfrak{so}(2)_l$ and $\mathfrak{so}(2)_r$ generators. In this case both the $\mathfrak{su}(2)_f$ and $\mathfrak{u}(1)_f$ act non-trivially  on the gravitational states in the $\bH^{\rho}_{2}\otimes \tX^{\rho}$ sector, while the left   R-symmetry $\mathfrak{u}(2)_R$ goes along for the ride to become the gravitational R-symmetry. 

For the $(3, 1)$ pair we have an $\N=4$ parent,
\be
\begin{array}{llllllllllllll}
\text{Parent} &\bV_{2}&\otimes& \tbV_{2}&=&\bG_{4}\oplus2\bV_4 \\[8pt]
\text{Big twin}&\bV_{1}\oplus\bC^{\rho}_{1}&\otimes& \tbV_{2}&=& \bG_{3}\oplus\bV_3 \\[8pt]
\text{Little twin} &\bV_{1}\oplus\bC^{\rho}_{1}&\otimes& \tA\oplus2\tX^{\rho}\oplus2\phi &=& \bG_{1}\oplus4\bV_{1}\oplus3\bC_{1} 
\end{array}
\ee
In this case, the left and right symmetries of the big twin factors are
\be
\mathfrak{so}(2)_l\oplus \mathfrak{u}(1)_R\oplus\mathfrak{u}(1)_f\quad\text{and}\quad \mathfrak{so}(2)_r\oplus \mathfrak{u}(2)_R,
\ee
where the right $\mathfrak{u}(1)_f$ is  a remnant  of the left $\nl=2$ R-symmetry. They sit inside  the $\N=3$ algebra as
\be
[\mathfrak{u}(2)_R\oplus\mathfrak{u}(1)_R]\oplus\mathfrak{u}(1)\subset \mathfrak{u}(3)_R\oplus\mathfrak{u}(1)_{\text{Isotropy}}
\ee
where as before the additional $\mathfrak{u}(1)$  is given by the difference of the $\mathfrak{so}(2)_l$ and $\mathfrak{so}(2)_r$ generators. Again, all gravitational states are uncharged under the  $\mathfrak{u}(1)_f$ as it only acts non-trivially on $\bC^{\rho}_1$. 

The left and right symmetries of the little twin factors are
\be
\mathfrak{so}(2)_l\oplus \mathfrak{u}(1)_R\oplus\mathfrak{u}(1)_f\quad\text{and}\quad \mathfrak{so}(2)_r\oplus \mathfrak{u}(2)_f,
\ee
where the right $\mathfrak{u}(2)_f$ is a remnant  of the right $\nr=2$ R-symmetry. They sit inside  the $\N=1$ algebra as
\be
\mathfrak{u}(1)_R\oplus[\mathfrak{u}(2)_f\oplus\mathfrak{u}(1)_f]\oplus\mathfrak{u}(1)\subset \mathfrak{u}(1)_R\oplus [\mathfrak{u}(3)\oplus\mathfrak{u}(1)]_{\text{Isotropy}}
\ee
The extra $\mathfrak{u}(1)$  is given by the difference of the $\mathfrak{so}(2)_l$ and $\mathfrak{so}(2)_r$ generators as in the other cases. In this case both the $\mathfrak{u}(2)_f$ and $\mathfrak{u}(1)_f$ act non-trivially  on the gravitational states in the $\bC^{\rho}_{1}\otimes 2\tX^{\rho}$ sector, while the left   R-symmetry $\mathfrak{u}(1)_R$ becomes  the gravitational R-symmetry. Note,  there is an additional global $\mathfrak{u}(1)$, which acts trivially on the scalars \cite{Ferrara:2010cw}.

Finally, for the $(2,1)$ pair we have an $\N=3$ parent that is not simply the product of pure $\N=2$ and $\N=1$ Yang-Mills,
\be
\begin{array}{llllllllllllll}\label{21pair}
\text{Parent} &\bV_{2}\oplus \bH_{2}^{\rho }&\otimes& \tbV_{1}\oplus \tbC_{1}^{\tilde{\rho}}&=&\bG_{3}\oplus3\bV_3 \\[8pt]
\text{Big twin}&\bV_{2}&\otimes& \tA \oplus 2\tX^{\tilde{\rho}}\oplus 2\tphi&=& \bG_{2}\oplus 3\bV_2 \\[8pt]
\text{Little twin} &\bV_{1}\oplus\bC^{\rho}_{1}&\otimes& \tA\oplus2\tX^{\rho}\oplus2\phi &=& \bG_{1}\oplus4\bV_{1}\oplus3\bC_{1} 
\end{array}
\ee
Note, to obtain the $\N^{+}_{-}=2$ twin we both decompose the right and truncate the left $\bH_{2}^{\rho }$ multiplet. Importantly, the $\N=3$ symmetry generated by $\bV_2\otimes\tbV_1$ alone is not enough to accommodate the required $\N^{+}_{-}=2$ R-symmetry plus isotropy, hence the need group theoretically for the two additional  $\bV_3$. Of course, they are also required  for the correct content.

Despite the fact that all three triplets $(3,2,1)$ are truncations of either the $\N=5$ or $\N=4$ parent,  symmetry considerations imply that in terms of Yang-Mills-matter factorisations considered here the $(3,1)$ and $(3,2)$ twins can only be accommodated by the $\N=4$ and $\N=5$ parents respectively.  

We note that the squaring approach provides a finer graining of the triplet than that of supergravity. From a supergravity perspective, the three sub twin pairs are truly degenerate in the triplet: treating $(3,2,1)$ as a triplet or as three pairs is equivalent. However, from the point of view of the double copy, the triplet degeneracy is partially resolved: not only do the $\susy=2$ and $\susy=3$ theories admit two different factorisations each, but these are in a $1-1$ correspondence with the sub-twin pair that they belong to. Thus, different factorisations uniquely lead to different sub pairs, and therefore different parents. As an example, given the $\susy=2$ theory along with its factorisation in \eqref{32pair}, one can determine that it belongs to the sub pair $(3,2)$, while the factorisation in \eqref{21pair} can only sit in the $(2,1)$ sub pair; this distinction cannot be actuated in supergravity. So far, only one factorisation of the $\susy=1$ has been found, such that the triplet is not yet fully resolved into three distinct sub pairs. We refer the reader to \cite{Anastasiou:2016alternatives} regarding the possibility of an alternative factorisation of the $\N=1$ theory.

Dimensionally reducing to $D=3$ we obtain a $(6, 4, 2)$ triplet with common scalar coset (with vectors dualised to scalars),
\be
\frac{\SU(4,2)}{\Un(4)\times\SU(2)}.
\ee

\subsection{Other twins}\label{othertwins}

There is an isolated pair of  twin supergravities that do not belong to the pyramid since the $\N_+$ twin is not a product of two pure super Yang-Mills theories: the $D=4, (4,1)$ pair consisting of pure $\N_+=4$ supergravity and $\N_-=1$ supergravity coupled to $n_V=6$ vector multiplets and a single chiral multiplet as discussed in \autoref{twins}. Recall, both have U-duality group $\SL(2, \R)\times \SO(6)$ and the common scalar manifold is $\SU(1,1)/\Un(1)$.  Note however, the pure $\N_+=4$ theory \emph{can} be considered as part of the pyramid if the $\N=0$ vector multiplet is included in the factors,
\be
\bV_4\otimes\tA=\bG_4,\qquad
\text{with} \qquad
[\mathfrak{so}(2)_l\oplus\mathfrak{su}(4)]\oplus[\mathfrak{so}(2)_r]\rightarrow\mathfrak{so}(2)_{st}\oplus\mathfrak{u}(4),
\ee
 where the spacetime helicity group $\mathfrak{so}(2)_{st}$ and the additional $\mathfrak{u}(1)$  are given by the sum and difference of the $\mathfrak{so}(2)_l$ and $\mathfrak{so}(2)_r$ generators, respectively.   The simplest approach\footnote{This is by no means  unique, there is for example an $\N=4$ parent. We leave the reader to explore the possibilities.} is to take $\N=5$ supergravity as the parent:
 \be
\begin{array}{llllllllllllll}
\text{Parent} &\bV_{4}&\otimes& \tbV_{1}&=&\bG_{5} \\[8pt]
\text{Big twin}&\bV_{4}&\otimes&\tA\oplus\tX^{\tilde{\rho}}&=& \bG_{4} \\[8pt]
\text{Little twin} &A\oplus4\chi^{\rho}\oplus6\phi&\otimes& \tbV_1&=& \bG_{1}\oplus6\bV_{1}\oplus\bC_{1} 
\end{array}
\ee
where, departing from the pyramid twins, we decompose either the left or right multiplet, but not both.  As usual the  R-symmetry of the $\N_+=4$ theory becomes the isotropy group of the $\N_-=1$ theory.

Finally, we recall that the $D=4, (2,1)$ twin pair appearing in  the pyramid admits a generalisation  to a two integer parameter  sequence of $(2,1)$ twins, as pointed out in  \cite{Duff:2010ss}. The $\N_+=2$   sequence is given by $\N=2$ supergravity minimally coupled to  $n_V$ vector multiplets and $n_H$ hyper multiplets. The $\N_-=1$ sequence is given by $\N=1$ supergravity  coupled to  $n_V+1$ vector multiplets and $n_C=n_V+2n_H$ chiral multiplets. The common  scalar coset is $\Un(1, n_V)\times \SU(2, n_H)/[\Un(1)\times \Un(n_V)\times\Un(2)\times\SU(n_H)]$. Arbitrary $n_V, n_H$ requires a  sequence of $\N=3$  parent supergravities coupled to   $n= n_V+ n_H$ vector multiplets. The $\N_\pm$  scalar cosets then embed into the parent $\N=3$  scalar coset,
\be\label{N3N2break}
\frac{\Un(1, n_V)\times \Un(2, n_H)}{\Un(1)\times\Un(n_V)\times\Un(2)\times\Un(n_H)}\subset\frac{\Un(3,n)}{\Un(3)\times\Un(n)}.
\ee
With two supersymmetric factors this is achieved by including fundamental matter from the outset,
\be
[\bV_2\oplus\bC_{2}^{\rho}]\otimes[\tbV_1\oplus m\tbC_{1}^{\rho}]=\bG_3 \oplus n\bV_3,\qquad
\ee
where $n=m+1$ and 
\be
[\mathfrak{so}(2)_l\oplus\mathfrak{u}(2)_R]\oplus[\mathfrak{so}(2)_r\oplus\mathfrak{u}(1)_R\oplus\mathfrak{u}(m)]\rightarrow\mathfrak{so}(2)_{st}\oplus[\mathfrak{u}(3)_R\oplus \mathfrak{u}(n)].
\ee

To obtain the $\N_+=2$ twin sequence as a truncation of the $\N=3$ parent sequence we decompose the right multiplet,
\be
\tbV_1\oplus m\tbC_{1}^{\rho} \longrightarrow \tA \oplus \tX \oplus m\tX^{\tilde{\rho}}\oplus m\tphi^{\tilde{\rho}},
\ee
where $\tphi$ is complex. We then further truncate $n_H+1=n-(n_V-1)$ of the spinors and $n_V-1=n-(n_H+1)$ of the scalars so that we break to
$
\mathfrak{u}(n_V-1)\oplus\mathfrak{u}(n_H)\subset\mathfrak{u}(m),
$
leaving
\be
[\bV_2\oplus\bC_{2}^{\rho}]\otimes[\tA \oplus   (n_V-1)\tX^{\tilde{\rho}}\oplus n_H\tphi^{\tilde{\rho}}]=\bG_2\oplus n_V\bV_2\oplus n_H \bH_2
\ee
with symmetries 
\be\label{syms1}
[\mathfrak{so}(2)_l\oplus\mathfrak{u}(2)_R]\oplus[\mathfrak{so}(2)_r\oplus\mathfrak{u}(1)\oplus\mathfrak{u}(n_V-1)\oplus\mathfrak{u}(n_H)]\rightarrow\mathfrak{so}(2)_{st}\oplus[\mathfrak{u}(2)_R\oplus\mathfrak{u}(1)\oplus \mathfrak{u}(n_V)\oplus \mathfrak{u}(n_H)]
\ee
in agreement with \eqref{N3N2break}.

To obtain the $\N_-=1$ twin sequence as a truncation of the $\N=3$ parent sequence we decompose both the left and right multiplets,
\be
\begin{split}
\bV_2\oplus \bC_{2}^{\rho} &\longrightarrow \bV_1\oplus \bC_{1} \oplus \bC_{1}^{\rho},\\
\tbV_1\oplus m\tbC_{1}^{\rho} &\longrightarrow \tA \oplus \tX \oplus m\tX^{\tilde{\rho}}\oplus m\tphi^{\tilde{\rho}},
\end{split}
\ee
where $\tphi$ is complex. Again, we  then further truncate both the left and right. On the left we only keep  $\tbC_{1}^{\rho}$, while on the right we remove  $n_H-1=n-(n_V+1)$ of the spinors and $n_V=n-n_H$ of the scalars so that we break to
$
\mathfrak{u}(1)\oplus\mathfrak{u}(n_V)\oplus\mathfrak{u}(n_H-1)\subset\mathfrak{u}(m),
$
where one of the right $\tX$ is a $\mathfrak{u}(n_V)$ singlet. This  yields
\be
[\bV_1\oplus\bC_{1}^{\rho}]\otimes[\tA \oplus \tX^{\tilde{\rho}}\oplus  n_V\tX^{\tilde{\rho}}\oplus (n_H-1)\tphi^{\tilde{\rho}}]=\bG_1\oplus (n_V+1)\bV_1\oplus (2n_H+n_V) \bC_1
\ee
with symmetries 
\be\label{syms2}
[\mathfrak{so}(2)_l\oplus\mathfrak{u}(1)_R]\oplus[\mathfrak{so}(2)_r\oplus\mathfrak{u}(1)\oplus\mathfrak{u}(n_V)\oplus\mathfrak{u}(n_H-1)]\rightarrow\mathfrak{so}(2)_{st}\oplus[\mathfrak{u}(1)_R\oplus\mathfrak{u}(2)\oplus \mathfrak{u}(n_V)\oplus \mathfrak{u}(n_H)]
\ee
in agreement with \eqref{N3N2break}. Note, in \eqref{syms1} and \eqref{syms2} $n_V$ and $n_H$ are interchanged precisely because $\bV_2\otimes \tA=\bG_2\oplus \bV_2$ generates an extra vector multiplet whereas $\bV_1\otimes \tA=\bG_1\oplus \bC_1$ generates an extra chiral multiplet.
 
 \section{Conclusions}\label{conclusions}
 
 We have shown that all non-maximal supergravities theories in $3\leq D\leq 6$ that are the product of two super Yang-Mills multiplets have a twin supergravity. Moreover, it has been shown that the parent supergravity and  its twins are all related through their Yang-Mills factorisations in a uniform manner. As far as we are aware the matter coupled $\N_-$ twins generated this way have not been  double-copy constructed   previously, so that we add to the already substantial list of supergravities admitting a Yang-Mills factorisation. 
 
 In the course of studying the twin pyramid it has become clear that the factorisation of matter coupled supergravity theories is not necessarily unique. Note, it had already been previously observed that supergravities can admit factorisations into alternative multiplets, for example $D=3, \N=16$ supergravity is the square of both $\N=8$ Yang-Mills theory and Bagger-Lambert-Gustavsson Chern-Simons theory \cite{Bargheer:2012gv}. In future work \cite{Anastasiou:2016alternatives} we will give a complete classification, using symmetry generating constraints, of all alternative factorisations and double-copy constructible theories under the assumption that the (super)gravity scalar manifold is symmetric. 
 
Considering the off-shell symmetries of the double-copy including fundamental matter multiplets  led us to introduce a bi-fundamental scalar field that couples to a bi-adjoint scalar field through a cubic interaction. Interestingly, it seems that bi-adjoint/fundamental scalar theory yields the zeroth-copy of Yang-Mills-matter amplitudes at tree-level, suggesting a generalisation of the spin 2,1,0 CHY formulae of \cite{Cachazo:2013iea} to include non-adjoint fields.

\section*{Acknowledgements} 
We are grateful to Zvi Bern, Marco Chiodaroli, Henrik Johansson and  Alexander Ochirov for useful discussions. The work of LB is supported by a Schr\"odinger Fellowship.  MJD is grateful to the Leverhulme Trust for an Emeritus Fellowship. LB would like to thank the Isaac Newton Institute for Mathematical Sciences, Cambridge, for support and hospitality during the programme Gravity, Twistors and Amplitudes where part of this work was done. AM would like to thank the School of Theoretical Physics, IAS Dublin,
for kind hospitality and inspiring environment. This work was  supported by the STFC under rolling grant ST/G000743/1.

\appendix

\section{Irreducible Riemannian globally symmetric K\"ahler and Quaternionic Manifolds}\label{MANIFOLDS}

Kahler manifolds:\\
\begin{align}
&\frac{\SU(p,q)}{\SU(p)\times{\SU(q)}\times{\Un(1)}}&&,&&\frac{\SO(p,2)}{\SO(p)\times{\Un(1)}}&&,&&\frac{\Sp(p,\R)}{\SU(p)\times\Un(1)}&&\\\nonumber
&\frac{\SO^*(2p)}{\SU(p)\times{\Un(1)}}&&,&&\frac{E_{6(-14)}}{\SO(10)\times{\Un(1)}}&&,&&\frac{E_{7(-25)}}{E_6\times{\Un(1)}}\\\nonumber
\end{align}

Quaternionic manifolds:\\
\begin{align}
&\frac{\SU(p,2)}{\SU(p)\times{\SU(2)}\times{\Un(1)}}&&,&&\frac{\SO(p,4)}{\SO(p)\times{\SO(4)}}&&,&&\frac{\USp(p,1)}{\USp(p)\times{\Un(1)}}&&\\\nonumber
&\frac{G_2}{\SU(2)\times{\SU(2)}}&&,&&\frac{F_{4(4)}}{\USp(3)\times{\Un(1)}}&&,&&\frac{E_{6(2)}}{\SU(6)\times{\SU(2)}}&&\\\nonumber
&\frac{E_{7(-5)}}{\SO(12)\times{\SU(2)}}&&,&&\frac{E_{8(-24)}}{E_7\times{\SU(2)}}
\end{align}

\section{Bi-adjoint coupled to bi-fundamental scalar theory: a 5-point example}\label{App5point}

\begin{figure}[h]
\includegraphics[width=0.95\textwidth]{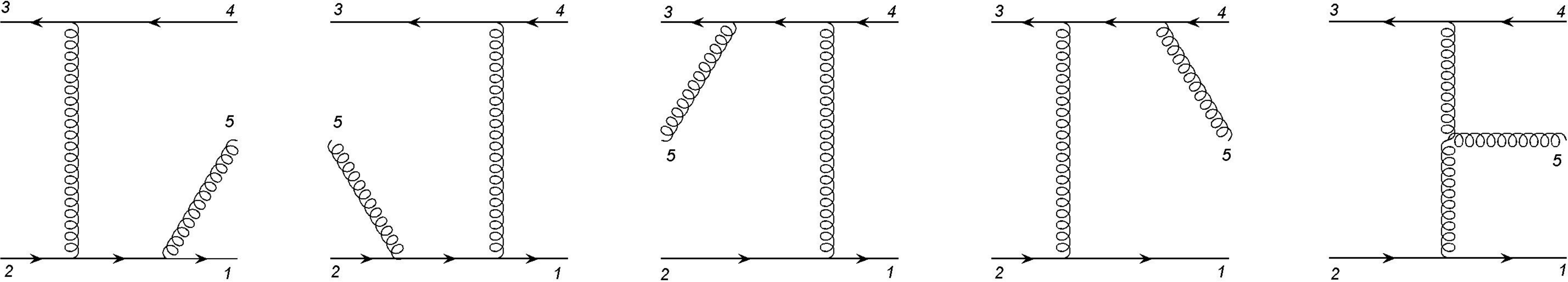}
\caption{Tree-level Feynman diagrams for the  5-point quark-antiquark-quark'-antiquark'-gluon interaction.}\label{5point_fig}
\end{figure}

In \autoref{twinsquare} we discussed how, starting with amplitudes for adjoint and fundamental fields, one can take the so-called ``zeroth-copy'' and generate amplitudes for a bi-adjoint scalar theory coupled to a bi-fundamental scalar. Here we give a 5-point example with two quark flavours, the double-copy of which was treated in \cite{Johansson:2014zca}. This nicely illustrates the point that the bi-adjoint/fundamental scalar theory can be embellished to capture flavour groups and other structural features appearing in the gauge theory. We start with the scattering amplitude of two quark-antiquark pairs, with distinct flavours, with a single gluon. The two flavours reduces the Feynman diagrams to the five presented in \autoref{5point_fig}.

\begin{figure}[h]
\includegraphics[width=0.95\textwidth]{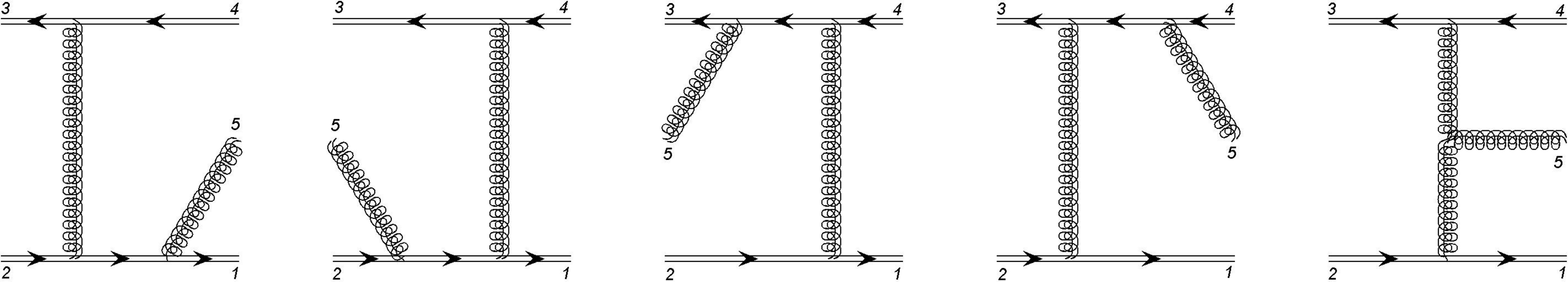}
\caption{Double-line tree-level Feynman diagrams for the  5-point (bi-fund.)-(bi-fund.)-(bi-fund'.)-(bi-fund'.)-(bi-adj.)~interaction for the scalar theory in \eqref{bifundscalar_flavour}. The curly (straight) double-lines represent the bi-adjoint (bi-fundamental) representation of the global $G\times \tG$ symmetry. Each diagram is the double-copy of the corresponding gluon-quark diagram shawn of its kinematic data.}\label{5point_double_fig}
\end{figure}

We use the labelling  $(1,\bar{u}_1(k_1),^i)$ and $(2,v_2(k_2),_j)$ for one of the quark-antiquark pairs, then $(3,\bar{u}_3(k_3),^k)$ and $(4,v_4(k_4),_l)$ for the other (possessing a different flavour) and finally $(5,\varepsilon^\mu_5(k_5),a)$ for the gluon. The colour and kinematic factors for these diagrams are then
\be
\begin{aligned}
c_1\times n_1&=-i\ [T^a]_i^{\ m}[T^b]_m^{\ j}[T^b]_k^{\ l}\quad&\times&\quad
\bar{u}_3\gamma^\nu v_4\ \bar{u}_1\varepsilon_5^\mu\gamma_\mu\gamma^\rho (k_1+k_5)_\rho\gamma_\nu v_2\\
c_2\times n_2&=-i\ [T^b]_i^{\ m}[T^a]_m^{\ j}[T^b]_k^{\ l}\quad&\times&\quad
\bar{u}_3\gamma^\nu v_4\ \bar{u}_1\gamma_\nu\gamma^\rho (-k_2-k_5)_\rho\varepsilon_5^\mu\gamma_\mu v_2 \\
c_3\times n_3&=-i\ [T^a]_k^{\ m}[T^b]_m^{\ l}[T^b]_i^{\ j}\quad&\times&\quad
\bar{u}_3\gamma_\mu\varepsilon_5^\mu\gamma^\rho(k_3+k_5)_\rho\gamma^\nu v_4\ 
\bar{u}_1\gamma_\nu v_2\\
c_4\times n_4&=-i\ [T^b]_k^{\ m}[T^a]_m^{\ l}[T^b]_i^{\ j}\quad&\times&\quad
\bar{u}_3\gamma^\nu\gamma^\rho(-k_4-k_5)_\rho\varepsilon_5^\mu\gamma_\mu v_4\ 
\bar{u}_1\gamma_\nu v_2\\
c_5\times n_5&=\quad \ \ f^{abc}[T^b]_i^{\ j}[T^c]_k^{\ l}\quad &\times&\quad 
\bar{u}_3\gamma^\nu v_4\ 
[\eta_{\mu\rho}(k_5-q)_\nu+\eta_{\rho\nu}(q-p)_\mu+
\eta_{\nu\mu}(p-k_5)_\rho]\varepsilon_5^\mu\  
\bar{u}_1\gamma^\rho v_2
\end{aligned} 
\ee
where, for the last diagram, $q=k_1+k_2$ and $p=k_3+k_4$. We know that these numerators automatically satisfy BCJ relations, so we can proceed as in the 4-point example and replace the kinematic factors with a second set of colour factors,
\be
\begin{aligned}
c_1\times \tilde{c}_1&=-i\ [T^a]_i^{\ m}[T^b]_m^{\ j}[T^b]_k^{\ l}\quad&\times&\quad  [\tilde T^{\tilde a}]_{\tilde i}^{\ \tilde m}[\tilde T^{\tilde b}]_{\tilde m}^{\ \tilde j}[\tilde T^{\tilde b}]_{\tilde k}^{\ \tilde l}\\
c_2\times \tilde{c}_2&=-i\ [T^b]_i^{\ m}[T^a]_m^{\ j}[T^b]_k^{\ l}\quad&\times&\quad   [\tilde T^{\tilde b}]_{\tilde i}^{\ \tilde m}[\tilde T^{\tilde a}]_{\tilde m}^{\ \tilde j}[\tilde T^{\tilde b}]_{\tilde k}^{\ \tilde l}\\
c_3\times \tilde{c}_3&=-i\ [T^a]_k^{\ m}[T^b]_m^{\ l}[T^b]_i^{\ j}\quad&\times&\quad   [\tilde T^{\tilde a}]_{\tilde k}^{\ \tilde m}[\tilde T^{\tilde b}]_{\tilde m}^{\ \tilde l}[\tilde T^{\tilde b}]_{\tilde i}^{\ \tilde j}\\
c_4\times \tilde{c}_4&=-i\ [T^b]_k^{\ m}[T^a]_m^{\ l}[T^b]_i^{\ j}\quad&\times&\quad   [\tilde T^{\tilde b}]_{\tilde k}^{\ \tilde m}[\tilde T^{\tilde a}]_{\tilde m}^{\ \tilde l}[\tilde T^{\tilde b}]_{\tilde i}^{\ \tilde j}\\
c_5\times \tilde{c}_5&=f^{abc}[T^b]_i^{\ j}[T^c]_k^{\ l}\quad&\times&\quad  
\tilde f^{\tilde a\tilde b\tilde c}[\tilde T^{\tilde b}]_{\tilde i}^{\ \tilde j}[\tilde T^{\tilde c}]_{\tilde k}^{\ \tilde l}
\end{aligned} 
\ee
and we see that these  reproduce the amplitude for a bi-adjoint scalar theory coupled to a pair of bi-fundamental scalars, with a non-trivial flavour group (see \autoref{5point_double_fig}). 
The cubic interactions for this theory are described by the Lagrangian 
\be\label{bifundscalar_flavour}
\mathcal{L}_{\text{bi-adj-fund}}=-\frac{1}{2}\partial_\mu \Phi_{a\tilde{b}} \partial^\mu \Phi^{a\tilde{b}}-\frac{1}{2}\partial_\mu \Phi_{i\tilde{i}}^\alpha \partial^\mu \Phi^{i\tilde{i}}_\alpha+\frac{g}{6}\left(f_{abc}\tilde{f}_{\tilde{a}\tilde{b}\tilde{c}}\Phi^{a\tilde{a}} \Phi^{b\tilde{b}} \Phi^{c\tilde{c}}  +i[T^{a}]_{i}{}^{j}[{\tilde{T}}^{\tilde{a}}]_{\tilde{i}}{}^{\tilde{j}}\Phi_{a\tilde{a}} \Phi^{i\tilde{i}}_\alpha \Phi_{j\tilde{j}}^\alpha\right),
\ee
where $\alpha$ denotes the representation of the flavour group, which the scalars have inherited from the original theory. We note that the replacement rules postulated in \eqref{repl1} and \eqref{repl2} still hold since the Feynman diagrams are already BCJ-duality respecting (i.e.~there are no four-point contact terms in this example).

%\bibliography{Ref_Library}
%\end{document}


\end{document}